\documentclass{elsart}
\usepackage{graphicx}
\usepackage{amssymb}
\usepackage{latexsym}
\usepackage{xcolor}

\begin{document}

\begin{frontmatter}
%
%
\title{Self-sustained current oscillations in the kinetic theory of semiconductor superlattices}
\author{E. Cebri\'an}
 \address[label1]{Departamento de Matem\'aticas y Computaci\'on. Universidad
de Burgos, 09001 Burgos, Spain.}
\author{L. L. Bonilla}
\address[label2]{G. Mill\'an Institute of Fluid Dynamics, Nanoscience and
Industrial Mathematics, Universidad Carlos III de Madrid, Avenida de la Universidad 30,
28911 Legan{\'e}s, Spain.}
\author{A. Carpio}
\address[label3]{Departamento de Matem\'atica Aplicada, Fac.\ Matem\'aticas,
 Universidad Complutense de Madrid, 28040 Madrid, Spain.}
\date{\today}
%
%
\newcommand{\N}{{\rm I}\!{\rm N}}
\newcommand{\R}{{\rm I}\!{\rm R}}
%
%
\begin{abstract}
We present the first numerical solutions of a kinetic theory description of
self-sustained current oscillations in n-doped semiconductor superlattices.
The governing equation is a single-miniband Boltzmann-Poisson transport equation
with a BGK (Bhatnagar-Gross-Krook) collision term. Appropriate boundary
conditions for the distribution function describe electron injection in the contact regions.
These conditions seamlessly become Ohm's law at the injecting contact and the
zero charge boundary condition at the receiving contact when integrated over
the wave vector. The time-dependent model is numerically solved for the
distribution function by using the deterministic Weighted Particle Method.
Numerical simulations are used to ascertain the convergence of the method.
The numerical results confirm the validity of the Chapman-Enskog perturbation method
used previously to derive generalized drift-diffusion equations for high electric fields because
they agree very well with numerical solutions thereof.
\end{abstract}

\end{frontmatter}

\section{Introduction}
When non-interacting electrons in the conduction band of a material are subject to a
constant electric field $E$, their positions should oscillate with a frequency proportional to
the electric field, $\omega_{B}=eEl/\hbar$, where $-e<0$, $\hbar$ and $l$ are the charge
of the electron, the Planck constant and the crystal period. These coherent Bloch oscillations
(BO) and the associated current were predicted by Zener in 1934 \cite{zener}.
Scattering limits the observability of BO: to observe them, their period should be smaller
than the scattering time $\tau$, so that $E>\hbar/(el\tau)$. In standard materials, the fields
required to observe BO are too large and therefore {\em damped} Bloch oscillations were not
found until 1992 in experiments with undoped semiconductor superlattices \cite{fel92},
which have much larger periods than natural crystals.

Semiconductor superlattices are artificial one-dimensional crystals formed by epitaxial growth
of layers belonging to two different semiconductors that have similar lattice constants
\cite{BGr05}. They were synthesized following Esaki and Tsu's idea that these artificial
crystals would be useful to realize BO or related high frequency oscillations \cite{esaki}.
The difference in the energy gaps of the component semiconductors makes the conduction
band of the superlattice to be a periodic succession of barriers and wells with typical
periods of several nanometers. The electronic spectrum of a superlattice (SL) consists of a
succession of minibands and minigaps generated by its periodicity. Tayloring the size of
barriers and wells and the negative doping density of the latter, it is possible to achieve SLs
with wide minibands and to populate only the lowest one. Electrons moving in this miniband
have energies that are periodic functions of their wave numbers and are scattered by 
phonons, impurities and other electrons. When an appropriate dc voltage
is held between the ends of one such SL with finitely many periods, it is possible to obtain
high-frequency self-sustained oscillations of the current through the structure \cite{BGr05}.
These oscillations are caused by repeated formation of electric field pulses at the injecting
contact of the SL that move forward and disappear at the receiving contact. Thus they are
transit-time oscillations whose frequency is inversely proportional to the SL length: they
are similar to the Gunn effect in bulk semiconductors \cite{kro72} and are different from BO.
These Gunn-type oscillations have been observed in experiments with GaAs/AlAs SL (and
with other SL based on III-V semiconductors) since 1996 and are the basis of fast oscillator
devices \cite{HGS96}. The connection between the existence of Gunn-type oscillations
and the suppression of Bloch oscillations is not yet well understood despite theoretical
and experimental efforts \cite{BGr05}.

Although mathematical models at the level of semiclassical kinetic theory go back
to the 1970s \cite{KSS-71}, their analysis has been based on simplified reduced ordinary
differential equations \cite{ISh87,IDS91} which typically ignore space charge effects.
Electron transport in a single miniband SL can be described by a kinetic equation coupled
to a Poisson equation approximately describing the electric potential due to the other electrons
\cite{BEP-03}. A simple kinetic equation \cite{ISh87} contains an energy-dissipating
collision term of Bhatnagar-Gross-Krook (BGK) type \cite{BGK-54} and a simple
energy-conserving (but momentum-dissipating) collision term. This model does not
include coupling to the Poisson equation. An important point is that the
dispersion relation between miniband energy and momentum is periodic because this
periodicity gives rise to a relation between electron drift velocity and electric field which has
a maximum value \cite{BGr05}. Then the drift velocity decreases as the field increases for
large field values (negative differential mobility) and this in turn causes the Gunn-type
self-sustained current oscillations (SSCO) for appropriate bias and contact boundary
conditions \cite{BGr05}. These features are absent in the more usual Boltzmann-Poisson
systems with parabolic band dispersion relations. 

Recently, Bonilla et al \cite{BEP-03} have derived a nonlinear
drift-diffusion equation from the KSS-BGK kinetic model coupled to the Poisson equation,
which we will call the BGK-Poisson system. They use a Chapman-Enskog
perturbation method in a particular limit in which the collision terms are of the same order as the
term containing the electric field and dominate all other terms in the kinetic equation. Then
stable SSCO are obtained by numerically solving the drift-diffusion equation with appropriate 
boundary and initial conditions \cite{BEP-03}. However, no one has solved
numerically the kinetic equation directly and shown that self-oscillations are among its
solutions or studied the relation between these solutions and those of the limiting
drift-diffusion equation. These are the problems tackled in the present paper and solving
them could be a step in more precise studies of stable current oscillations in superlattices
and other low dimensional solid state systems.

We solve the BGK-Poisson kinetic equation model by means of a deterministic weighted
particle method that has been used in the past to solve Boltzmann equations with
non-periodic energy band dispersion relations \cite{NIC-DEG-POU-88}. Particle methods
(see a recent one in \cite{yossi}) are appropriate to study our system of equations because
their solutions may present large gradients: the electric field pulses obtained by simulating the
approximate drift-diffusion equations have a smooth leading front but a steep trailing back
front \cite{BGr05}. The present work paves the way to numerically solving interesting
problems in nanoelectronics and spintronics that are described by related quantum kinetic
equations with more than one miniband \cite{BBA}.

\section{The Model}
Our model for electron transport in a single miniband SL is a
Boltzmann-Poisson system with BGK collision term \cite{BGK-54}
plus appropriate boundary and initial conditions. The governing
equations are:
\begin{equation}
\label{1}
\partial_{t}f + v(k) \partial_{x} f+\frac{eF}{\hbar}\partial_{k}f =
-\nu_{en} \left(f-f^{FD}\right) - \frac{\nu_{imp}}{2} [f-f(x,-k,t)],
\end{equation}
\begin{equation}
\label{2} \epsilon \partial_{x}^{2}V=\frac{e}{l} \left(n-N_{D}\right),
\quad F=\partial_{x}V,
\end{equation}
\begin{equation}
\label{3} n=\frac{l}{2\pi}\int_{-\frac{\pi}{l}}^{\frac{\pi}{l}}
f(x,k,t) \, dk=\frac{l}{2\pi}\int_{-\frac{\pi}{l}}^{\frac{\pi}{l}}
f^{FD}(k;\mu(n)) \, dk,
\end{equation}
\begin{equation}
\label{4}
f^{FD}(k;\mu)=\frac{m^{*}k_{B}T}{\pi\hbar^{2}}\ln\left[1+\exp\left(\frac{\mu-
\varepsilon(k)}{k_{B}T}\right)\right],
\end{equation}
with $x \in [0,L]$ and $f$ periodic in $k$ with period $2\pi/l$.
Here $l$, $L=Nl$, $N$, $\epsilon$, $f$, $n$, $N_{D}$, $k_{B}$, $T$, $V$, $-F$,
$m^{*}$ and $-e<0$ are the SL period, the SL length, the number of SL periods,
the dielectric constant, the
one-particle distribution function, the 2D electron density, the
2D doping density, the Boltzmann constant, the lattice
temperature, the electric potential, the electric field, the effective mass of the
electron, and the electron charge, respectively. We shall describe boundary and initial
conditions later.

The first term in the right hand side of Eq.\ (\ref{1}) represents
energy relaxation towards a 1D effective Fermi-Dirac distribution
$f^{FD}(k;\mu(n))$ \cite{BEP-03} (local equilibrium) due to e.g.\
phonon scattering. $\nu_{en}$ is the collision frequency, taken as constant for
simplicity. Here, $\mu(n)$ is the chemical potential that is a function of $n$ resulting from 
solving equation (\ref{3}) when (\ref{4}) is substituted in it. A similar BGK model
with a Boltzmann local distribution function was proposed by Ignatov and Shashkin
\cite{ISh87,IDS91}. The second term in the right hand side of Eq.\
(\ref{1}) accounts for impurity elastic collisions with the
constant collision frequency $\nu_{imp}$, which conserve energy
but dissipate momentum \cite{KSS-71,BEP-03,BGr05}. Transfer of lateral
momentum due to impurity scattering \cite{GER-93} is ignored in
this model. We assume the simple tight-binding miniband dispersion
relation,
\begin{equation}
\varepsilon(k) = \frac{\Delta}{2}\left(1-\cos kl\right)\Longrightarrow
v(k)=\frac{1}{\hbar}\,\frac{d\varepsilon}{dk}=\frac{l\Delta}{2\hbar}\,\sin kl,
\label{disprel}
\end{equation}
where $\Delta$ is the miniband width. The exact and Fermi-Dirac
distribution functions, $f$ and $f^{FD}$, have the same electron
density $n$, according to (\ref{3}). The latter equation is solved
for the chemical potential $\mu$ in terms of $n$, which yields the
function $\mu(n)$. When (\ref{1}) is integrated over $k$, we
obtain the charge continuity equation,
\begin{eqnarray}
\label{continuity} \partial_{t}n + \frac{l}{e}\partial_{x}J_{n}=0,
\quad\mbox{with}\\
 J_{n}=\frac{e}{2\pi}\int_{-\pi/l}^{\pi/l} v(k)\, f(x,k,t)\, dk, \label{Jn}
\end{eqnarray}
where $J_{n}$ is the electron current density.

\subsection{Voltage bias condition}
Using the Poisson equation (\ref{2}) to eliminate
$n$, we obtain the following form of Amp\`ere's law:
\begin{eqnarray}
\label{ampere} \epsilon\,\partial_{t}F + J_{n}=J(t),
\end{eqnarray}
where $J(t)$ is the total current density. The total current density can be obtained from
the voltage bias condition:
\begin{equation}
\Phi(t)\equiv \frac{1}{L}\int_{0}^L F(x,t)\, dx = \phi,
\label{bias}
\end{equation}
where $\Phi(t) L$ is the voltage between the two contacts at the end of the SL and $\Phi(t)$
is an average field. For dc voltage bias, $\Phi(t)$ is a fixed constant $\phi$. If we integrate
(\ref{ampere}) over $x$ and use (\ref{bias}), we obtain
\begin{equation}
 J(t)=\frac{1}{L}\int_{0}^L J_{n}(x,t)\, dx. \label{J}
\end{equation}

\subsection{Boundary conditions}
The boundary conditions give the distribution function $f$ on the
contacts at $x=0$ and $x=L$ through the distribution function
inside the semiconductor. For fixed $|k|$, there are two possible
characteristic curves at a point $(x,t)$: one for $k>0$ and
another one for $k<0$. With $k<0$ the characteristic curve for
$x\to 0+$ and $t>0$ is given by the initial condition whereas it
is given by the distribution function at the contact ($x=0$) if
$k>0$. Then, for $x=0$ we need to specify the distribution
function at the contact for $k>0$, $f^+$, whereas for $x=L$ we
need to specify the distribution function at the contact for
$k<0$, $f^-$. Instead of inventing a theory for injecting and
collecting contacts, we use a top-down approach proposed in Ref.\
\cite{BGr05}: we know that the following boundary conditions
appropriately describe current self-oscillations in the
drift-diffusion equation for the electric field,
\begin{eqnarray}
&& J_{n}(0,t) =\sigma\, F(0,t), \label{bc1}\\
&& n(L,t)=N_{D}, \label{bc2}
\end{eqnarray}
where $\sigma>0$ is the constant contact conductivity and the left hand side of Eq.\
(\ref{bc1}) is the electron current density. We will use boundary conditions for $f$
such that they become (\ref{bc1}) and (\ref{bc2}) when we integrate them according to
the definitions (\ref{3}) and (\ref{Jn}) of $n$ and $J_{n}$ respectively:
\begin{equation}
\label{6} f^{+} = \frac{2 \pi \hbar \sigma F}{e \Delta}-
\frac{f^{(0)}}{\int_{0}^{\frac{\pi}{l}} v(k)f^{(0)} \, dk}
\int_{-\frac{\pi}{l}}^{0} v(k)f^{-} \, dk,
\end{equation}
for $x=0$, and
\begin{equation}
\label{7} f^{-} =\frac{f^{(0)}}{(l/2\pi)\int_{-\frac{\pi}{l}}^{0}
f^{(0)} \, dk} \left( N_{D}-\frac{l}{2\pi}\int_{0}^{\frac{\pi}{l}}
f^{+} \, dk \right),
\end{equation}
for $x=L$. Note that the integral over $k$ of (\ref{6}) times $e\, v(k)/(2\pi)$ yields
(\ref{bc1}) and the integral over $k$ of (\ref{7}) times $l/(2\pi)$ yields (\ref{bc2}). In
these equations, $f^{(0)}$ is the leading order approximation for the distribution function in
the Chapman-Enskog method \cite{BEP-03,BGr05}:
\begin{equation}
\label{5} f^{(0)}(k;n,F) = \sum_{j=-\infty}^{\infty} f_{j}^{(0)}
\exp\left( \imath jkl \right),
\end{equation}
where
\[
f_{j}^{(0)}=\frac{1-\imath j \varphi /
\tau_{e}}{1+j^2\varphi^2}f_{j}^{FD}
\]
\[
f_{j}^{FD}=\frac{l}{\pi}\int_{0}^{\frac{\pi}{l}} f^{FD} \cos(jkl)
\, dk
\]
\[
\varphi=\frac{F}{F_{M}} \qquad
F_{M}=\frac{\hbar\sqrt{\nu_{en}(\nu_{en}+\nu_{imp})}}{el} \qquad
\tau_{e}=\sqrt{1+\frac{\nu_{imp}}{\nu_{en}}}.
\]
Eq.\ (\ref{5}) is the solution of (\ref{1}) when we drop the $x$ and $t$ derivatives of $f$
(see \cite{BEP-03}).

If we use the electric potential $V$ instead of the field $F=\partial V/\partial x$ (recall
that the true electric field is $-F$), the following boundary conditions for $V$ are compatible
with (\ref{bias}):
\begin{eqnarray}
\label{biasV} V(0,t)=0, \quad V(L,t)=\phi L=\int_{0}^L F(x,t)\, dx.
\end{eqnarray}

\subsection{Initial condition}
We select (\ref{5}) as our initial condition for the distribution
function. The initial electric field is assumed to be constant,
$F(x,0)=\phi$, where $\phi$ is the average field. If we start from
other initial conditions, the evolution of the current and other
magnitudes are similar to those presented here after about 0.3 ps.

Recapitulating, the equations governing our model are (\ref{1}) - (\ref{4}) for the
unknowns $f$ and $V$ with initial condition (\ref{5}) and boundary conditions (\ref{6}),
(\ref{7}) and (\ref{biasV}). If we use the field $F$ instead of the electric potential
$V$, the voltage bias condition (\ref{bias}) for $F$ replaces (\ref{biasV}).

\section{Nondimensional equations}
\label{sec:nondim}
We use the scales defined in Table 1 to nondimensionalize the
Boltzmann-BGK-Poisson kinetic equations. These scales are based on
the hyperbolic scaling explained in Ref.\ \cite{BEP-03}

\[\hskip -8mm
\begin{array}{c}
\begin{array}{|c|c|c|c|c|c|c|}
  \hline
  F & V& k & x & t & \mu & n,\, f \\
  \hline
  F_{M} & F_{M}x_{0}& 1/l & x_{0} & t_{0} & k_{B}T & N_{D} \\
  \hline
  F^{a}=\frac{F}{F_M} &V^{a}=\frac{V}{F_{M}x_0}& k^{a}=kl & x^{a}=\frac{x}
  {x_0} & t^{a}=\frac{t}{t_0} &\mu^{a}=\frac{\mu}{k_{B}T} & n^{a}=\frac{n}{
  N_D}, \, f^{a}=\frac{f}{N_D} \\
  \hline
\end{array}
\\
{\rm Table \, 1: \, Hyperbolic \, scaling.}
\end{array}
\]

\[
x_{0}=\frac{\epsilon F_{M} l}{e N_{D}},
\qquad t_{0}=\frac{x_{0}}{v_{M}} ,
\qquad
v_{M}=\frac{\Delta l \Im_{1}
\left(\widetilde{M}\right)}{4\hbar\tau_{e}\Im_{0}
\left(\widetilde{M}\right)},
\]
\[
\Im_{j}\left(\widetilde{M}\right)=\frac{1}{2\pi}\int_{-\pi}^{\pi}
\cos (j k^{a})\, \ln\left[1+\exp\left(\widetilde{M}-\delta+\delta
\cos(k^{a})\right)\right]\, dk^{a},
\]
where $\widetilde{M}$ verifies
\[
1=\frac{\alpha}{2\pi}\int_{-\pi}^{\pi} \ln\left[1+\exp\left(\widetilde{M}
-\delta+\delta\cos(k^{a})\right)\right]\, dk^{a},
\]
with
\[
\alpha=\frac{m^{*}k_{B}T}{\pi\hbar^{2}N_{D}}.
\]
Numerical values for these parameters will be given in Section \ref{sec:num}.
Equations (\ref{1}) - (\ref{4}) have the following
nondimensional form
\begin{eqnarray}\nonumber
&&\partial_{t^{a}}f^{a} + \frac{\Delta l}{2\hbar v_{M}}\sin(k^{a})
\partial_{x^{a}} f^{a}+\frac{\tau_{e}}{\eta}F^{a}\partial_{k^{a}}f^{a}=
\\
\label{8}&&\quad\quad
\frac{1}{\eta} \left[f^{FDa}(k^{a};\mu^{a}(n^{a})) -\left( 1+\frac{
\nu_{imp}}{2\nu_{en}} \right) f^{a} + \frac{\nu_{imp}}{2
\nu_{en}} f^{a}(x^{a},-k^{a},t^{a}) \right],
\end{eqnarray}
\begin{equation}
\label{9} \partial_{x^{a}}^{2}V^{a} = \partial_{x^{a}} F^{a}
=n^{a}-1
\end{equation}
\begin{equation}
\label{10} n^{a}=\frac{1}{2\pi}\int_{-\pi}^{\pi}
f^{a}(x^{a},k^{a},t^{a}) \, dk^{a}=\frac{1}{2\pi}\int_{-\pi}^{\pi}
f^{FDa}(k^{a};\mu^{a}(n^{a})) \, dk^{a}
\end{equation}
\begin{equation}
\label{11} f^{FDa}(k^{a};\mu^{a})=\alpha
\ln\left[1+\exp\left(\mu^{a}-\delta+\delta
\cos(k^{a})\right)\right]
\end{equation}
\[
\eta=\frac{v_{M}}{\nu_{en} x_{0}} \qquad \delta=\frac{\Delta}{2
k_{B} T}.
\]

The dimensionless boundary conditions are, for $x^{a}=0$:
\begin{equation}
\label{12} f^{a+} = \beta F^{a}  - \frac{f^{a(0)}}{\int_0^\pi \sin
\left( {k^a } \right) f^{a(0)} \,dk^{a}} \int_{- \pi}^{0} \sin
\left( k^{a} \right) f^{a-} \,dk^a
\end{equation}
with
\[ \beta  = \frac{{2\pi \hbar \sigma F_M }} {{e\Delta N_D }}
\]
and for $x^{a}=L/x_{0}$:
\begin{equation}
\label{13} f^{a-}= \frac{f^{a(0)}}{(1/(2\pi))\int_{-
\pi}^{0}f^{a(0)} \,dk^a } \left( 1 - \frac{1}{2\pi}\int_{0}^{\pi}
f^{a+} \, dk^a \right).
\end{equation}

The boundary conditions for the electric potential $V^{a}$ are
\begin{eqnarray}
\label{Vbc} V^{a}(0,t^{a})=0, \quad
V^{a}(L^{a},t^{a})=\phi^{a}L^{a}\equiv
\frac{\phi}{F_{M}}\frac{L}{x_{0}}.
\end{eqnarray}

The dimensionless initial condition is
\begin{equation}
\label{dic} f^{a(0)}(k^{a};n^{a}) = \sum_{j=-\infty}^{\infty}
\exp\left( \imath jk^{a} \right) \frac{1- \imath j F^{a} /
\tau_{e}}{1+j^2\left( F^{a} \right)^2}f_{j}^{FDa}(n^{a})
\end{equation}
\[
f_{j}^{FDa}(n^{a})=\frac{1}{\pi}\int_{0}^{\pi} f^{FDa}(k^{a};\mu^{a}
(n^{a}))\, \cos(jk^{a})\, dk^{a}
\]
with $x^{a} \in [0,L^{a}=L/x_{0}]$ and $f^{a}$ periodic in $k^{a}$
with period $2\pi$.

Besides the electron current density, $J_{n}$, it is convenient to calculate the average
energy $E$ (and its nondimensional version, $E^{a}$), defined as $E^{a}=E/(k_{B}T)$:
\begin{equation}
\label{energa}
E^{a}= \frac{\int_{-\pi/l}^{\pi/l}\varepsilon(k) f(x,k,t)
dk}{k_{B}T\int_{-\pi/l}^{\pi/l} f(x,k,t) dk} =\delta\, \frac{\int_{-\pi}^{\pi}
(1-\cos k^{a}) f^{a}(x^{a},k^{a},t^{a})
dk^{a}}{\int_{-\pi}^{\pi} f^{a}(x^{a},k^{a},t^{a}) \, dk^{a}}.
\end{equation}

From now on we drop the superscript $a$.

\section{The Deterministic Weighted Particle Method}

The most widely used numerical method used for solving Boltzmann
equations is the Monte-Carlo Method \cite{REG-83}. This stochastic
method yields data with a lot of numerical noise. The {\em
deterministic} Weighted Particle Method (WPM) is an interesting
alternative because it yields the distribution function (and
therefore its moments: electron density, average energy and
current density) at each time during the transient regimes with
much less noise than the Monte Carlo simulation; cf.\
\cite{NIC-DEG-POU-88,DEL-MUS-92,CE-MU-94} (a numerical analysis of
WPM can be found in \cite{DEG-NIC-89} and  in \cite{issa} for the
special case of the BGK equation of gas dynamics).

The WPM relies on a particle description of
the distribution function, which means that $f(x,k,t)$ is written
as a sum of delta functions 
\[
f(x,k,t) \approx \sum_{i=1}^{N}   \omega_{i}
 f_{i}(t) \delta(x-x_{i}(t)) \otimes \delta(k-k_{i}(t))
\]
where $\omega_{i}$, $f_{i}(t)$, $x_{i}(t)$ and $k_{i}(t)$ are,
respectively, the (constant) control volume, the weight, the
position and the wave vector of the $i$th particle. $N$ is the
number of numerical particles.

In the WPM, the motion of particles is governed by collisionless
dynamics, whereas the collisions are accounted for by the
variation of weights. Large gradients in the solution profile
arise from appropriate particles acquiring large weights, not by
accumulating many particles in the large gradient regions. The
evolution of the particles is determined by their positions and
wave vectors which are the characteristic curves of the convective
part of the equation. Their equations are:
\begin{equation}
\frac{d}{dt}k_{i}(t)  = \frac{\tau_{e}}{\eta}F_{i}(t), \quad
\frac{d}{dt}x_{i}(t) = \frac{\Delta l}{2\hbar v_{M}} \sin \left(
k_{i}(t) \right) \label{ecarac}
\end{equation}
where $F_{i}(t)=F(x_{i}(t),t)$ denotes the electric field at the instantaneous position of
the $i$-th particle.

The evolution of the weight $f_{i}(t)$ is given by the ordinary differential equation:
\begin{equation}
\frac{d}{dt}f_{i}(t) = \frac{1}{\eta} \left[- \left( 1 + \frac{\nu
_{imp}}{2\nu _{en}} \right)f_{i}(t)  + \frac{\nu_{imp}}{2\nu
_{en}}f \left(x_{i}(t),-k_{i}(t),t \right) + f^{FD}_{i}(t) \right]
\label{edist}
\end{equation}
with $f^{FD}_{i}(t)$ the Fermi-Dirac distribution evaluated for the $i$-th particle.

The system of ordinary differential equations (\ref{ecarac}) -
(\ref{edist}) is now solved by using a modified (semi-implicit)
Euler method:
\begin{equation}
f_{i}^{n}  = f_{i}^{n-1} + dt \frac{1}{\eta} \left[- \left( 1 +
\frac{\nu_{imp}}{2\nu _{en}} \right)f_{i}^{n-1}  + \frac{\nu
_{imp}} {2\nu _{en}}\hat{f}_{i}^{n-1} + f_{i}^{FD,n-1} \right]
\label{eulerw}
\end{equation}
with $\hat{f}_{i}^{n-1}=f(x_{i}^{n-1},-k_{i}^{n-1},t^{n-1}),$
\begin{equation}
k_{i}^{n} = k_{i}^{n-1} + dt \frac{\tau_{e}}{\eta}F_{i}^{n-1},
\label{eulerk}
\end{equation}
\begin{equation}
x_{i}^{n} = x_{i}^{n-1} + dt \frac{\Delta l}{2\hbar v_{M}} \sin
\left( k_{i}^{n} \right). \label{eulerx}
\end{equation}
For stability reasons, we use $k_{i}^n$ to update $x_{i}^n$. The standard Euler
method would use $k_{i}^{n-1}$ to update $x_{i}^n$ but this would require
using unpractically small time steps to have a stable scheme. The same problem
appears when we employ explicit Runge-Kutta or multi-step methods. To select
the initial positions and wave vectors in the modified Euler method, we build
a grid in the domain  $\left[ 0,L \right]\times \left[ -\pi,\pi \right]$
and choose the values $(x_{i}^0,k_{i}^0)$ as the cell centers. The weights
$f_{i}^0$ are then chosen according to (\ref{dic}).

The boundary conditions are taken into account as follows:
\begin{itemize}
\item If $k_{i}^{n} > \pi$, we set $k_{i}^{n} = k_{i}^{n} - 2\pi$.
If $k_{i}^{n} < -\pi$, we set $k_{i}^{n} = k_{i}^{n} + 2\pi.$
\item If $x_{i}^{n} > L$, we set $x_{i}^{n} = x_{i}^{n} - L$ and
$f_{i}^{n-1} = f_{i}^{+}$. If $x_{i}^{n} < 0$, we set $x_{i}^{n} =
x_{i}^{n} + L$ and $f_{i}^{n-1} = f_{i}^{-}.$ Here $f_{i}^{+}$ and
$f_{i}^{-}$ are calculated by discretization of the integrals in
(\ref{12}) and (\ref{13}) using the composite Simpson's rule on an equally
spaced mesh $K_{m'}$ with step $\Delta k.$
\end{itemize}

To calculate $x_{i}$, $k_{i}$ and $f_{i}$ at the next time step
$t^{n+1}$, we need to update the electric field and the
Fermi-Dirac distribution in the equations for the particles.
According to Eqs.\ (\ref{2}) and (\ref{3}), this updating requires
an interpolation procedure to generate an approximation of the
distribution function on a regular mesh $X_{m}$, $K_{m'}$ which is
then used to approximate the electric field and the chemical
potential. To approximate the values of the distribution function
over the mesh, $f_{m,m'}^n$, we use  the following weighted mean 
of its values for the particles, $f_{i}^{n}$:
\begin{equation}
f_{m,m'}^{n}  = \frac{{\sum\limits_{i = 1}^N {f_i^{n} W_{m,m'}^i }
}}{{\sum\limits_{i = 1}^N {W_{m,m'}^i } }} \label{fdm}
\end{equation}
where
\[
W_{m,m'}^i  = \max \left\{ {0,1 - \frac{{\left| {X_{m}  - x_i^{n}
} \right|}}{{\Delta x}}} \right\} \cdot \max \left\{ {0,1 -
\frac{{\left| {K_{m'}  - k_i^{n} } \right|}}{{\Delta k}}} \right\}
\]
and $\Delta x$ and $\Delta k$ are the spatial and wave vector steps.

An approximation for the density (\ref{10}) and average energy
(\ref{energa}) at the mesh points, $n\left( X_{m},t^{n}
\right)\approx n_{m}^{n}$ and $\left(k_{B} T \right)^{-1} E\left(
X_{m},t^{n} \right)\approx \left(k_{B} T \right)^{-1} E_{m}^{n}$,
are obtained using the composite Simpson's rule and the interpolated values of
the distribution function on the mesh.

We calculate the nondimensional chemical potential $\mu$ by
using a Newton-Raphson iterative scheme to solve equations
(\ref{10}) and (\ref{11}):
\begin{equation}
\mu_p  = \mu _{p - 1}  - \frac{{g\left( {\mu _{p - 1} } \right)}}
{{g'\left( {\mu _{p - 1} } \right)}} \label{newton}
\end{equation}
with
\[
g\left( {\mu } \right) =  n  - \frac{\alpha}{2\pi} \int_{ - \pi }^\pi {\ln
\left[ {1 + \exp \left( {\mu  - \delta  + \delta \cos \left( {k}
\right)} \right)} \right]\,dk}
\]
 \[
 g'\left( {\mu } \right) =  - \frac{\alpha}{2\pi} \int_{ -
\pi }^\pi  {\frac{{\exp \left( {\mu  - \delta  + \delta \cos
\left( {k } \right)} \right)}} {{1 + \exp \left( {\mu  - \delta +
\delta \cos \left( {k} \right)} \right)}}\,dk }.
\]
The initial guess for $\mu$ is obtained by plotting $g(\mu)$ and selecting
a value near its zero. $g(\mu)$ and $g'(\mu)$ are evaluated using the composite
Simpson's rule. Once we know the chemical potential $\mu$, Eq.\
(\ref{11}) provides the Fermi-Dirac distribution function at mesh
points, $f^{FD} \left( K_{m'}; \mu(n_{m}^{n}) \right)$, which is
then interpolated to get the Fermi-Dirac weight function for
the particles, $f^{FD,n}_i$:
\begin{eqnarray}
f_{i}^{FD,n} &=& \left( \frac{X_{m+1}-x_{i}^{n}}{\Delta x}
\right)\left( \frac{K_{m'+1}-k_{i}^{n}}{\Delta k} \right) f^{FD}
\left( K_{m'}; \mu(n_{m}^{n}) \right)\nonumber
\\
&+& \left( \frac{x_{i}^{n}-X_{m}}{\Delta x} \right)\left(
\frac{K_{m'+1}-k_{i}^{n}}{\Delta k} \right) f^{FD} \left( K_{m'};
\mu(n_{m+1}^{n}) \right)\nonumber
\\
&+& \left( \frac{X_{m+1}-x_{i}^{n}}{\Delta x} \right)\left(
\frac{k_{i}^{n}-K_{m'}}{\Delta k} \right) f^{FD} \left( K_{m'+1};
\mu(n_{m}^{n}) \right)\nonumber
\\
&+&\left( \frac{x_{i}^{n}-X_{m}}{\Delta x} \right)\left(
\frac{k_{i}^{n}-K_{m'}}{\Delta k} \right) f^{FD} \left( K_{m'+1};
\mu(n_{m+1}^{n}) \right),\label{FDparticle}
\end{eqnarray}
provided the particle $i$ is in $\left[
X_{m},X_{m+1}\right]\times\left[ K_{m'},K_{m'+1}\right]$.

To compute the electric field at time $t^n$, we use finite differences to discretize the
Poisson equation on the grid $X_{m}:$
\begin{eqnarray}
&& \frac{ V_{m+1}^{n} - 2 V_{m}^{n} + V_{m-1}^{n} }{ \left(\Delta
x\right)^2} = n_{m}^{n} - 1,
\label{epoisson1} \\
&& F_{m}^{n} = \frac{ V_{m+1}^{n} - V_{m-1}^{n} }{ 2\, \Delta x}.
\label{epoisson2}
\end{eqnarray}
Here $V\left( 0,t^{n} \right)=0$ and $V\left( L,t^{n} \right)=\phi
L$ as indicated by (\ref{Vbc}). $V_{m}^{n}$ and $F_{m}^{n}$
denote our approximations of $V\left( X_{m},t^{n} \right)$ and
$F\left( X_{m},t^{n} \right)$ on the equally spaced mesh $X_{m}$.
Finally, the electric field is interpolated at the location of the
particle $i$
\begin{equation}
F_{i}^{n} = \left( \frac{X_{m+1}-x_{i}^{n}}{\Delta x} \right)
F_{m}^{n} + \left( \frac{x_{i}^{n}-X_{m}}{\Delta x} \right)
F_{m+1}^{n},Ê\label{Fparticle}
\end{equation}
provided the particle $i$ is in $\left[X_{m},X_{m+1}\right]$.

The total current density $J$ is given by Eq.\ (\ref{J}), whose nondimensional version is
\begin{equation}
J(t) = \frac{\varsigma}{L}\int_{0}^{L} \left[ \int_{-\pi}^{\pi}
\sin(k) f(x,k,t) \, dk \right] \, dx, \label{Jadim}
\end{equation}
in which
\[
\varsigma = \frac{l\Delta}{4\pi\hbar v_{M}}.
\]
We use the composite Simpson rule to approximate $J(t^{n})$.

Summarizing, at each time step $t^n$:
\begin{enumerate}
\item Calculate the boundary conditions (\ref{12}) and (\ref{13}) with data at
time $t^{n-1}$.
    \item Compute $f_{i}^n$, $k_{i}^n$ and $x_{i}^n$ according to
(\ref{eulerw}), (\ref{eulerk}) and (\ref{eulerx}), respectively, by
using their values at $t^{n-1}$.
    \item Evaluate the distribution function $f_{m,m'}^n$ at the mesh
points $(X_{m},K_{m'})$ by the weighted mean (\ref{fdm}).
\item Compute the electron density (\ref{10}) and nondimensional average energy (\ref{energa})
at the mesh points.
    \item Calculate the chemical potential (\ref{newton}) and compute the Fermi-Dirac
    distribution (\ref{11}) at the mesh points.
    \item Interpolate the Fermi-Dirac distribution (\ref{11}) at the mesh points to obtain
    the Fermi-Dirac weight function $f^{FD,n}_i$ according to (\ref{FDparticle}).
    \item Compute the electric field at the mesh points
by solving the finite difference discretization of the Poisson
equation, (\ref{epoisson1}) and (\ref{epoisson2}) and interpolate at the particles according to
(\ref{Fparticle}).
    \item Calculate the current by evaluation of (\ref{Jadim}) using the composite Simpson's rule.
\end{enumerate}

We have observed that the costlier processes are 2 (computation of $f_{i}^n$ using
(\ref{eulerw})) and 6 (computation of the Fermi-Dirac weight function $f_i^{FD,n}$):
these two processes take about 50\% of the overall computation time and they are
equally costly. After these processes, 3, 5 and 7 have the largest computational cost (each 
takes between 10\% and 19\% of the overall computation time). 

\section{Numerical results}
\label{sec:num}
We have used the parameter values
of \cite{ESC-BON-06}. Numerical solutions of the nonlinear
drift-diffusion equation derived from the Boltzmann-BGK model show
that there is a stable stationary state for voltage bias below a
certain threshold. Above this critical voltage, stable
self-sustained oscillations of the current appear. These
oscillations are due to the periodic generation of electric field
pulses at the injecting contact and their motion towards the
receiving contact. We have observed the same phenomena in our
numerical solutions of the Boltzmann-BGK kinetic equations.
Firstly, we present a typical case of self-sustained current
oscillations accompanied by the motion and recycling of an
electric field dipole wave, corresponding to a 157-period $3.64$
nm $GaAs$/$0.93$ nm $AlAs$ SL at 14 K, with $\Delta = 72$ meV,
$N_{D} = 4.57 \times 10^{10}$ cm$^{-2}$, $\nu_{imp} = 2 \nu_{en} =
18 \times 10^{12}$ Hz and dimensionless dc average field $\phi =1$
\cite{ESC-BON-06}. The constant conductivity is $2.5\, \Omega$
cm$^{-1}$ and the effective mass is $m^{*} = (0.067d_{W} +
0.15d_{B})m_{0}/l$, where $m_{0} = 9.109534 \times 10^{-31}$ kg is
the electron rest mass. Using these numerical values, the scales of 
space, time, velocity, electric field and dimensionless chemical potential defined in Section
\ref{sec:nondim} take on the following values: 
$$x_0= 15.92\,\mbox{nm},\, t_0=0.23\,\mbox{ps}, \, v_M=68.33\,\mbox{km/s},
\, F_M= 22.45\,\mbox{kV/cm},
\, \widetilde{M}=7.11.$$

\begin{figure}
\begin{center}
\includegraphics*[width=12.5cm]{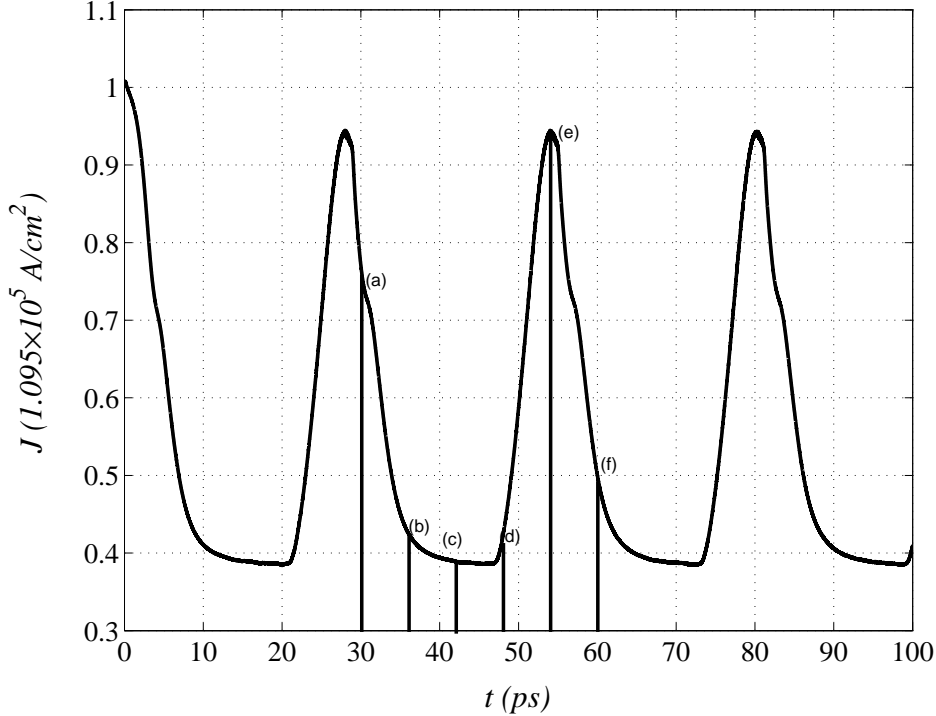}
\caption{Total current density versus time plot exhibiting self-sustained oscillations.
Units are written in parentheses. The oscillation period is 24 ps and the ratio between
the maximum and the minimum current is 2.6. At the times marked (a) - (f)
within one oscillation period (30, 36, 42, 48, 54 and 60 ps, respectively), we shall 
depict the electric field profile, the electron density profile, the distribution function and
the density plots thereof in Figures \ref{fig2}, \ref{fig7},\,Ê\ref{fig6} and \ref{fig6.2}, 
respectively.} 
\label{fig1}
\end{center}
\end{figure}

\begin{figure}
\begin{center}
\includegraphics*[width=12.5cm]{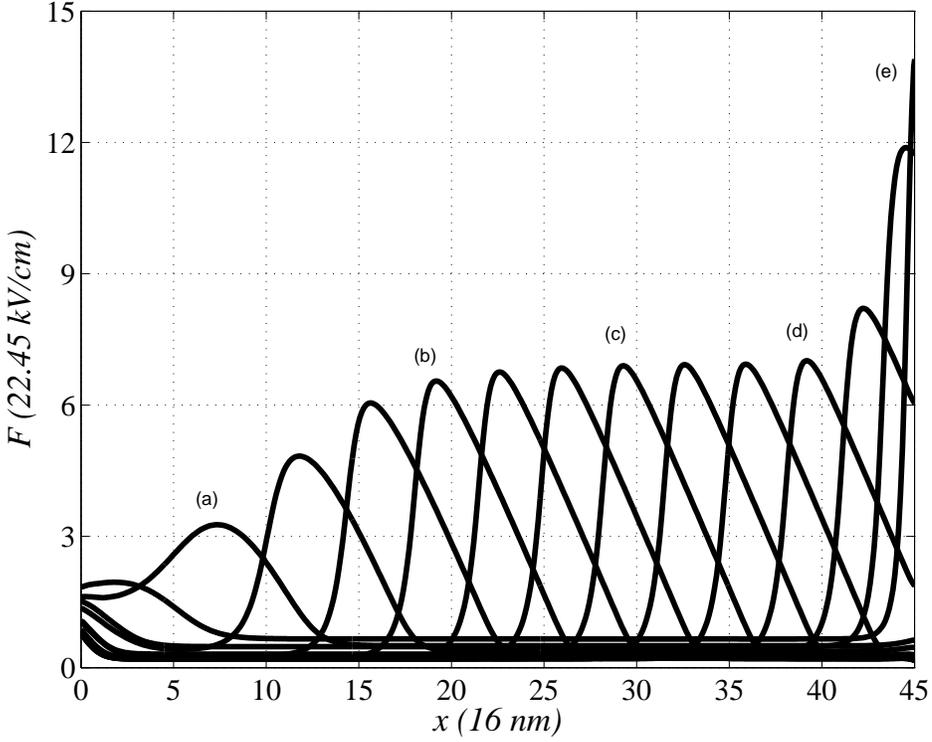}
\caption{Electric field versus position at different times within
one period of the oscillation. Far from the contacts, at time (c), the 
electric field pulse is 320 nm wide and 139 kV/cm tall. Thus it occupies 
about 45\% of the SL extension. At time (e), the electric field has a maximum 
value of 312 kV/cm. }
\label{fig2}
\end{center}
\end{figure}

\begin{figure}
\begin{center}
(a) \includegraphics*[width=6cm]{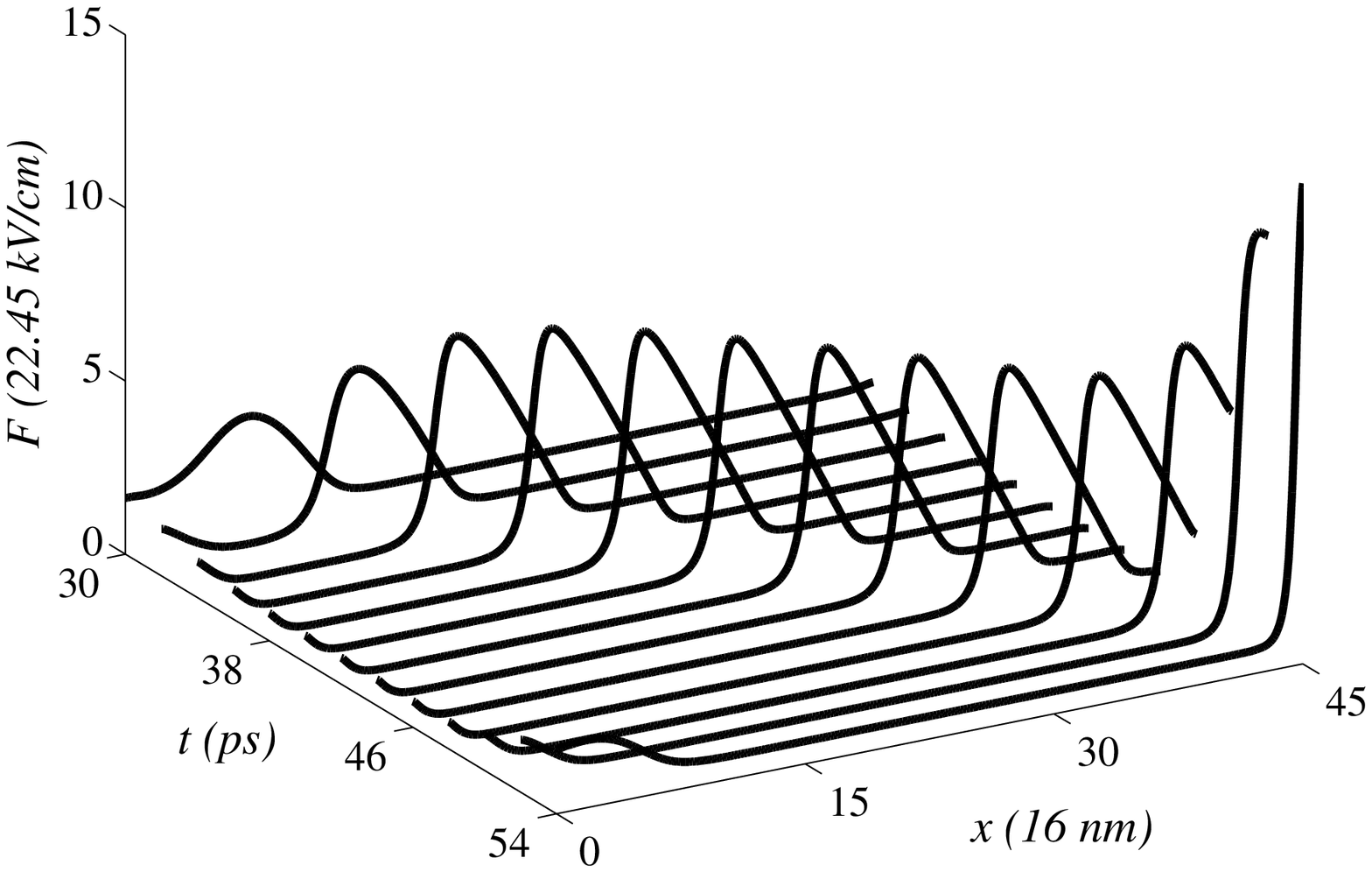}  (b)
\includegraphics*[width=6cm]{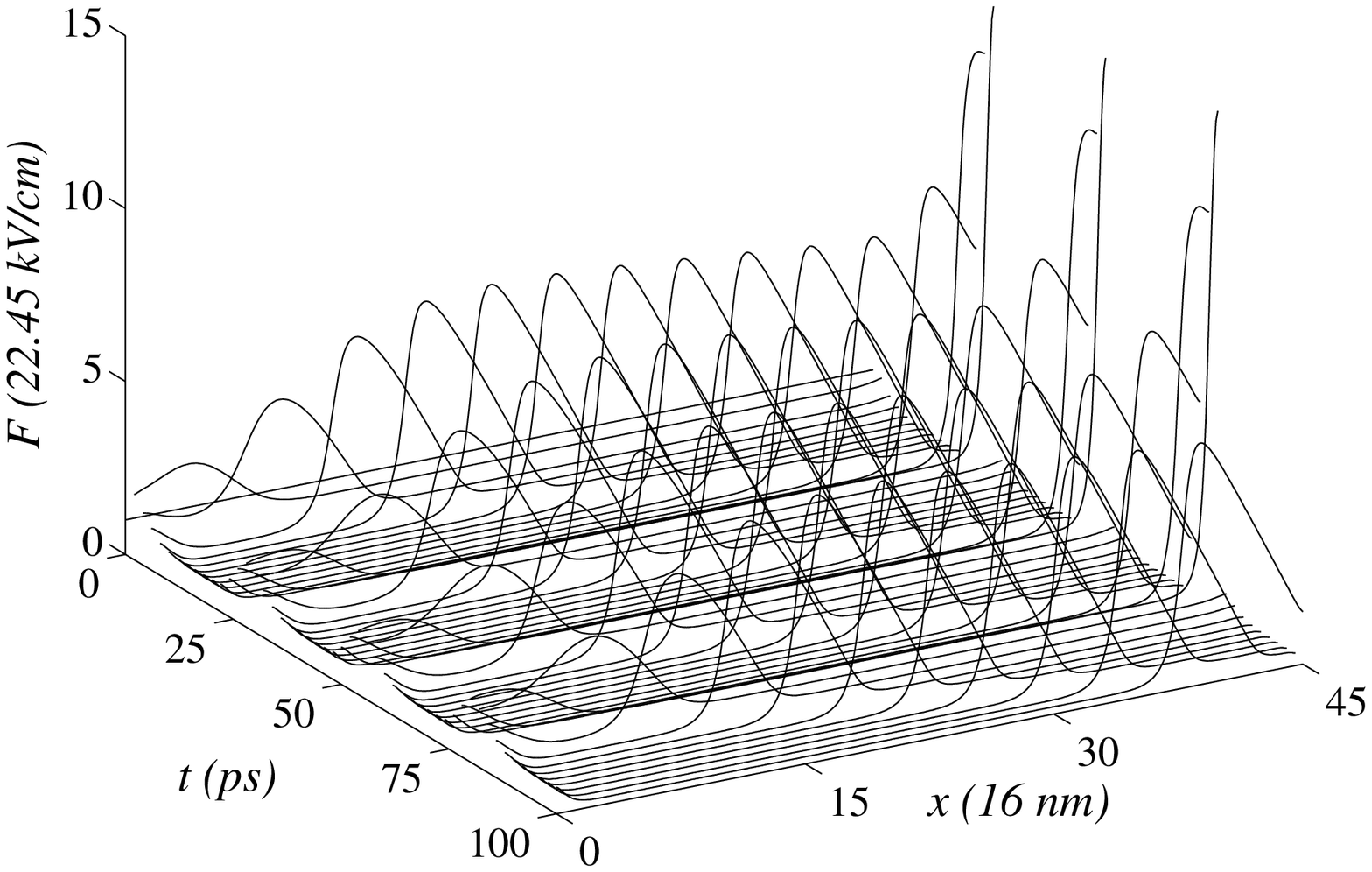} \caption{Evolution of
the electric field $F(x,t)$ during (a) one period and (b) several periods of the
self-oscillation.}\label{fig3}
\end{center}
\end{figure}

For these parameter values, we consider 140800 particles and a
mesh of 440 grid points for $x$ and 80 points for $k$. The time
step ($d t$) is $0.002$ ps. Figure \ref{fig1} shows the
self-oscillations of the current, and Figure \ref{fig2} the
corresponding electric field pulse at different times. We observe how the
electric field pulses are periodically created at the injecting
contact $x=0$, move to the end of the SL and disappear at the
receiving contact. In Fig.\ \ref{fig2}, we have depicted the field
profiles at the times marked (a) - (e) in Fig.Ê\ref{fig1}. We
observe that the total current density reaches its maximum value
when the electric field pulse is about to disappear at the
collector. The electric field as a function of time and position
is shown in Figure \ref{fig3}, both during one oscillation period
in Fig.\ \ref{fig3}(a) and during several periods in Fig.
\ref{fig3}(b). The ratio from the maximum to the minimum current in Fig.Ê\ref{fig1} 
is 2.6 whereas the same ratio calculated by solving the drift-diffusion
equation derived in \cite{ESC-BON-06} is 2.1 (cf.\ dashed line in Fig.\ 1(a) of 
\cite{ESC-BON-06}). Measured in units of $t_0$ (which has a different 
numerical value in \cite{ESC-BON-06}), the oscillation period is
104.3 in Fig.\ \ref{fig1} whereas the drift-diffusion equation yields 113.8. 
Comparing Fig.\ 1(b) of \cite{ESC-BON-06} with our Fig.\ \ref{fig2}, we find that
at the time (c) the solution of the BGK-Poisson equation produces a pulse far
from the contacts which is $11\, x_0$ wide and $7\, F_M$ tall whereas the drift-diffusion
equation yields a similar pulse which is $10.7\, x_0$ wide and $6.8\, F_M$ tall
(cf.\ dashed line in Fig.\ 1(b) of \cite{ESC-BON-06}). Thus the agreement 
between the simulation of the BGK-Poisson system and that of the drift-diffusion
equation is very good considering  the approximations made
in the derivation of the latter from the former. 

\begin{figure}
\begin{center}
(a) \includegraphics*[width=6cm]{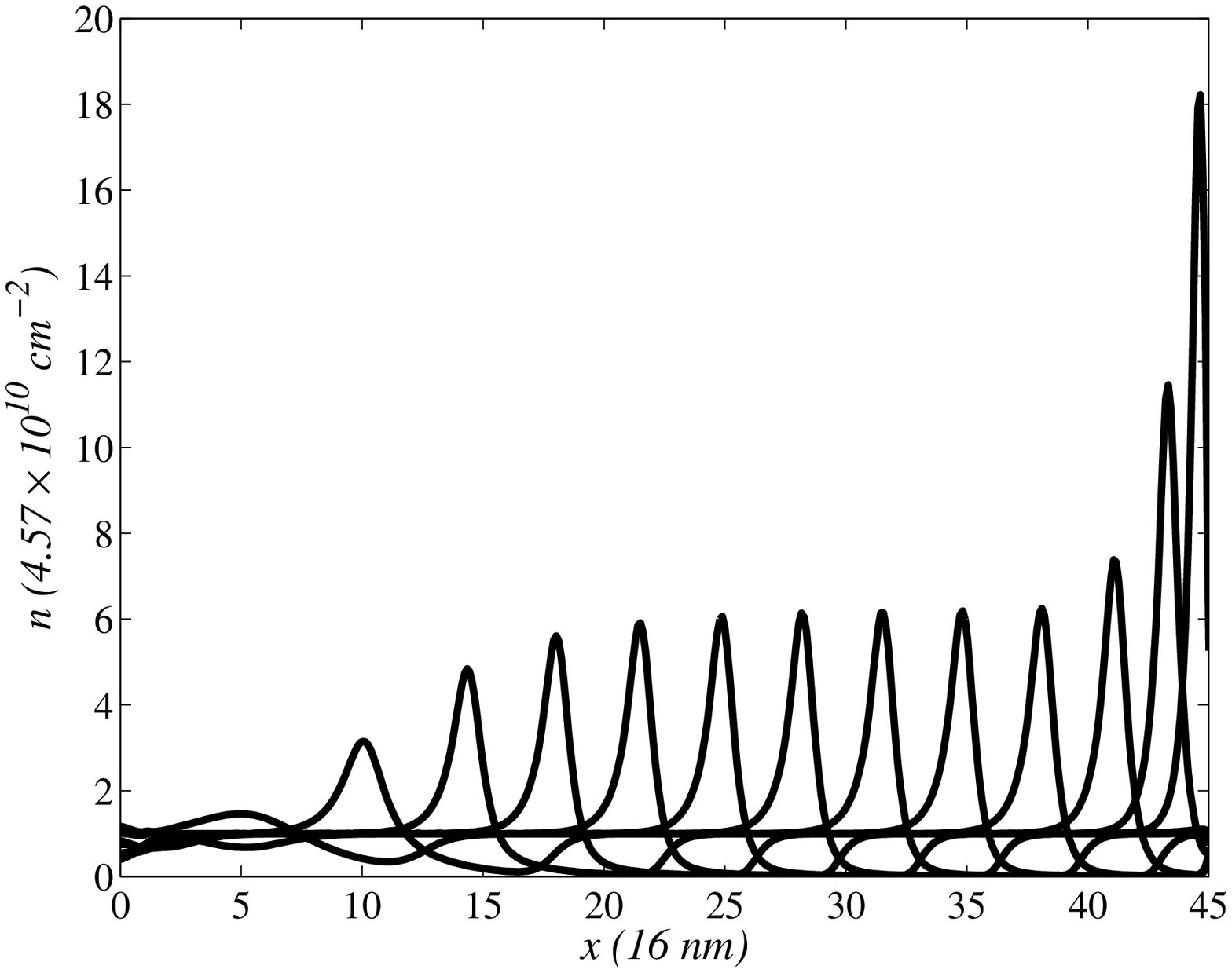} (b)
\includegraphics*[width=6cm]{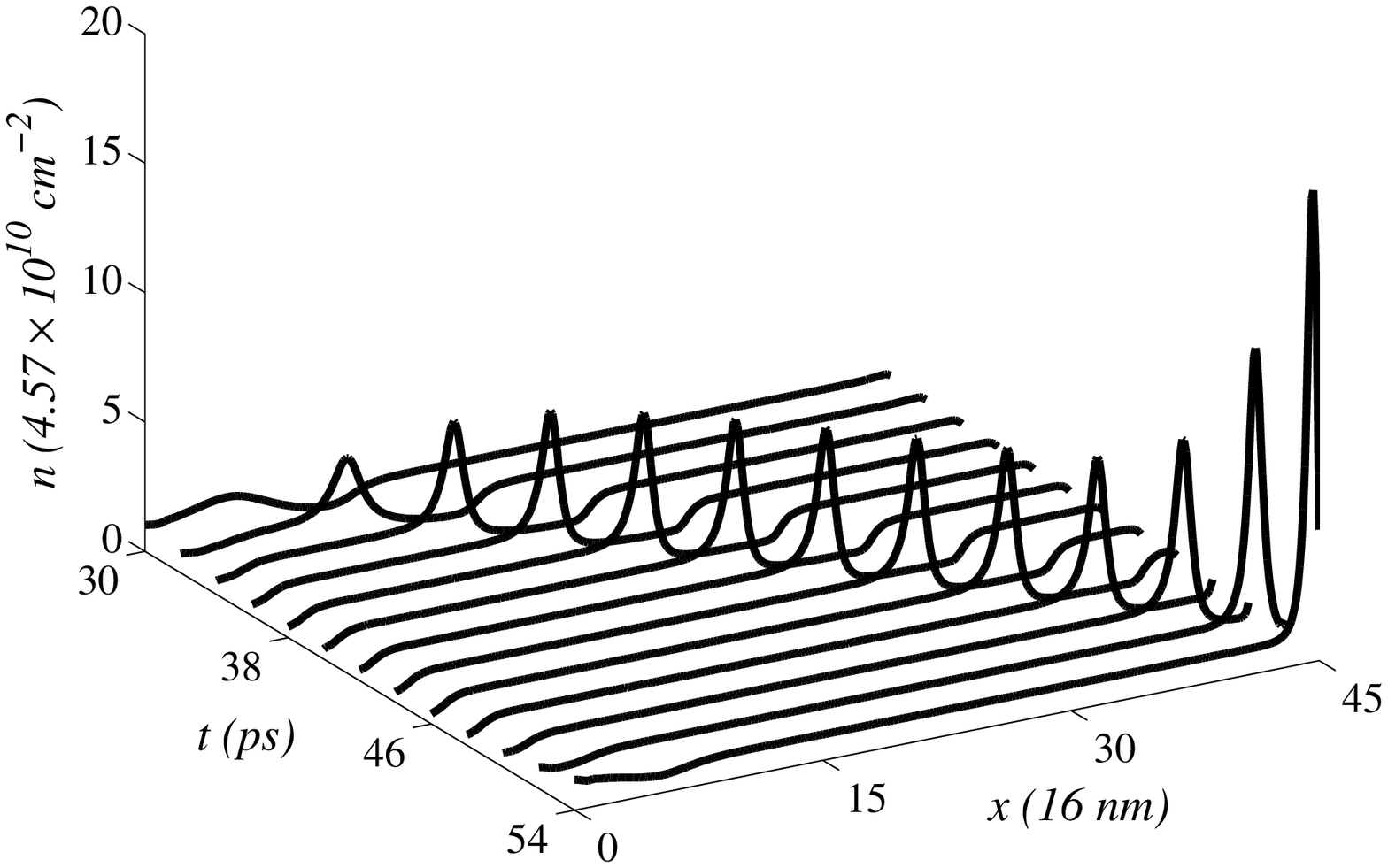} \caption{Electron
density profiles during one oscillation period.}\label{fig4}
\end{center}
\end{figure}

Figure~\ref{fig4} shows the dimensionless electron density. We see
the profile during several times belonging to one oscillation
period as a function of position in Fig.\ \ref{fig4}(a) and as a
function of the time and position in Fig.\ \ref{fig4}(b).
The electron density profile corresponding to an
electric field pulse is that of a traveling dipole wave such that $n>1$ behind the
peak of the electric field and $0<n<1$ ahead of the
peak. Comparison with Fig.\ \ref{fig3}(a) shows that the local
maximum of the electron density is reached somewhat later than the
peak of the electric field pulse.

\begin{figure}
\begin{center}
(a) \includegraphics*[width=6cm]{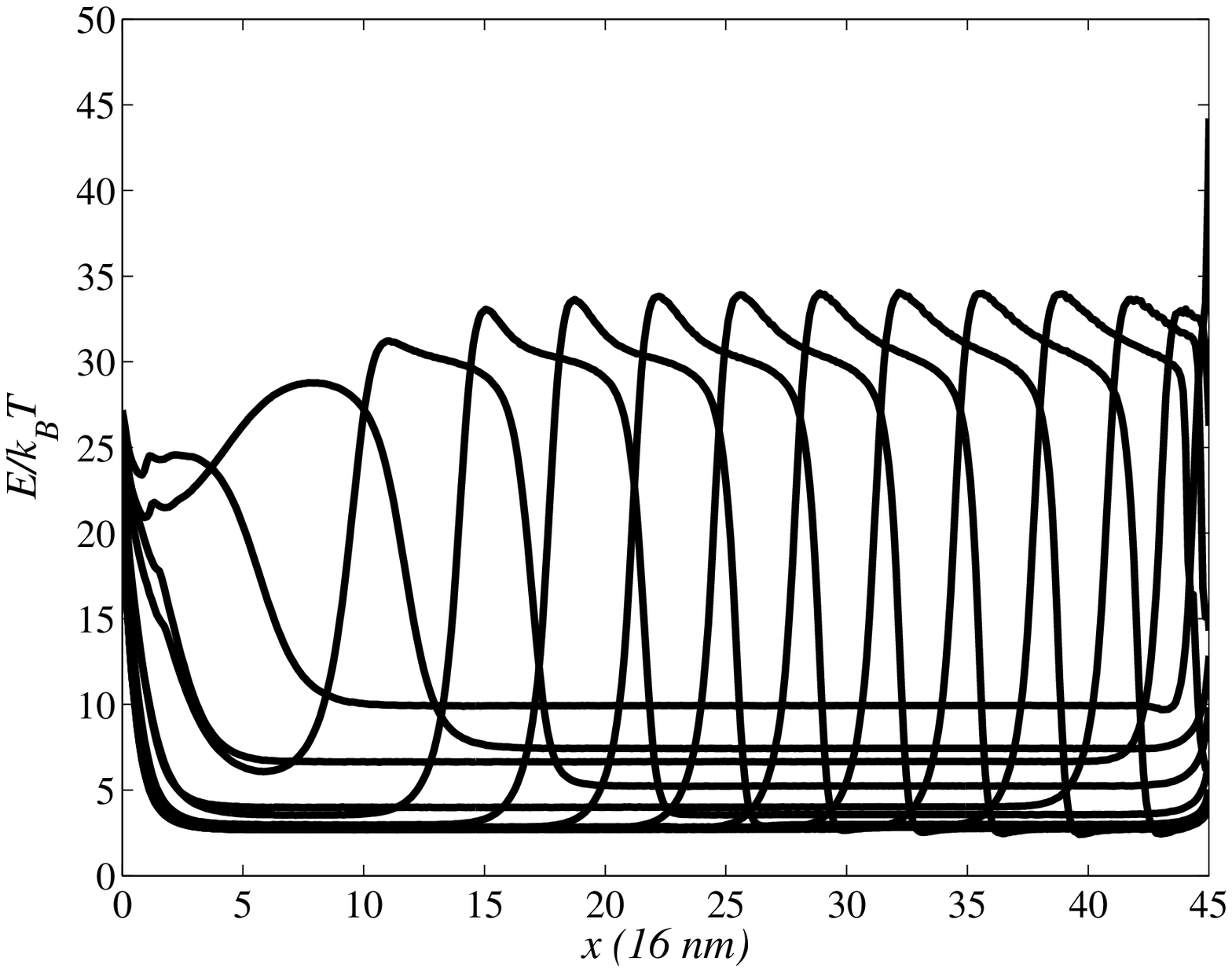} \hspace{3.mm} (b)
\includegraphics*[width=6cm]{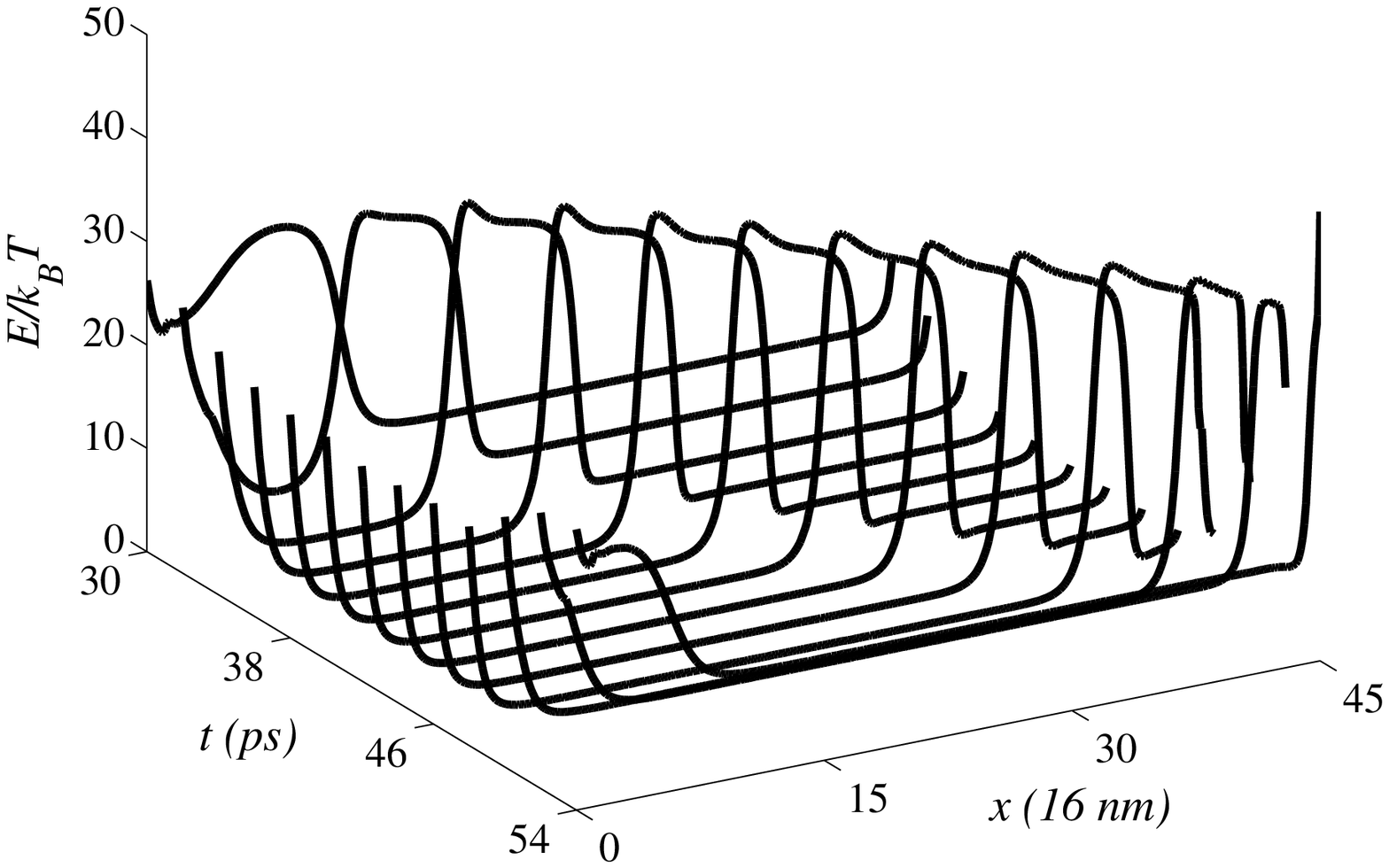}
\caption{Nondimensional average energy profiles, $E/(k_{B}T),$ at different
times of one oscillation period.}\label{fig5}
\end{center}
\end{figure}

Figure~\ref{fig5}(a) depicts the nondimensional average energy,
$E/(k_{B}T)$ as a function of distance at different instants of
one oscillation period. The average energy profile is pulse-like. Its local maximum
is always quite close to the peak of the electric field during each oscillation period.
 Fig.\ \ref{fig5}(b) shows the average energy profile as a function of position and time
during one oscillation period.

\begin{figure}
\begin{center}
\includegraphics*[width=12.5cm]{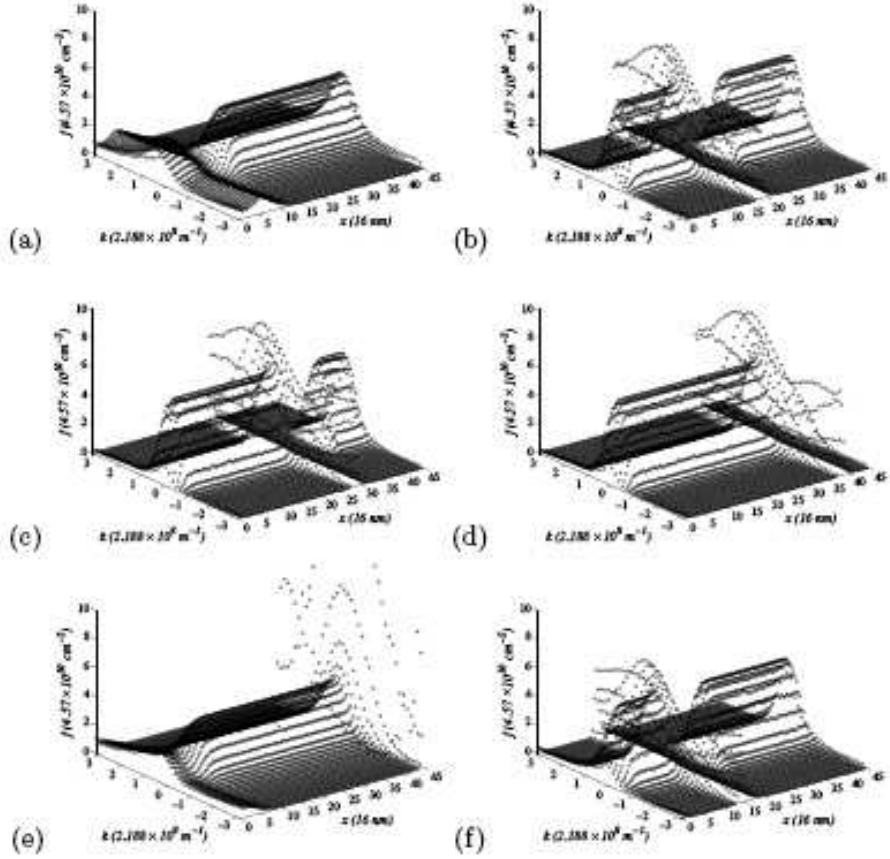}
\caption{(a) - (f) Distribution
function versus position and wave vector at the different times of one
oscillation period as marked in Fig.\ \ref{fig1}.}\label{fig6}
\end{center}
\end{figure}

\begin{figure}
\begin{center}
\includegraphics*[width=12.5cm]{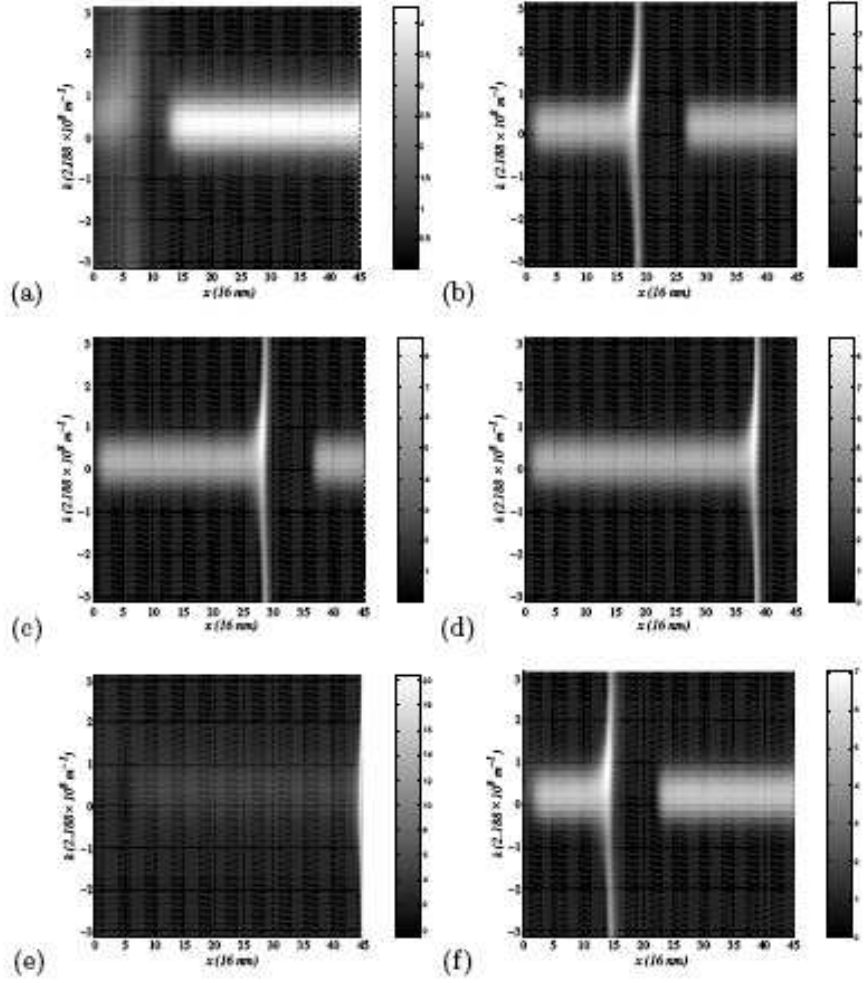}
\caption{(a) - (f) Density plots of the distribution
function versus position and wave vector at the different times of one
oscillation period as marked in Fig.\ \ref{fig1}.}\label{fig6.2}
\end{center}
\end{figure}

\begin{figure}
\begin{center}
(a) \includegraphics*[width=6cm]{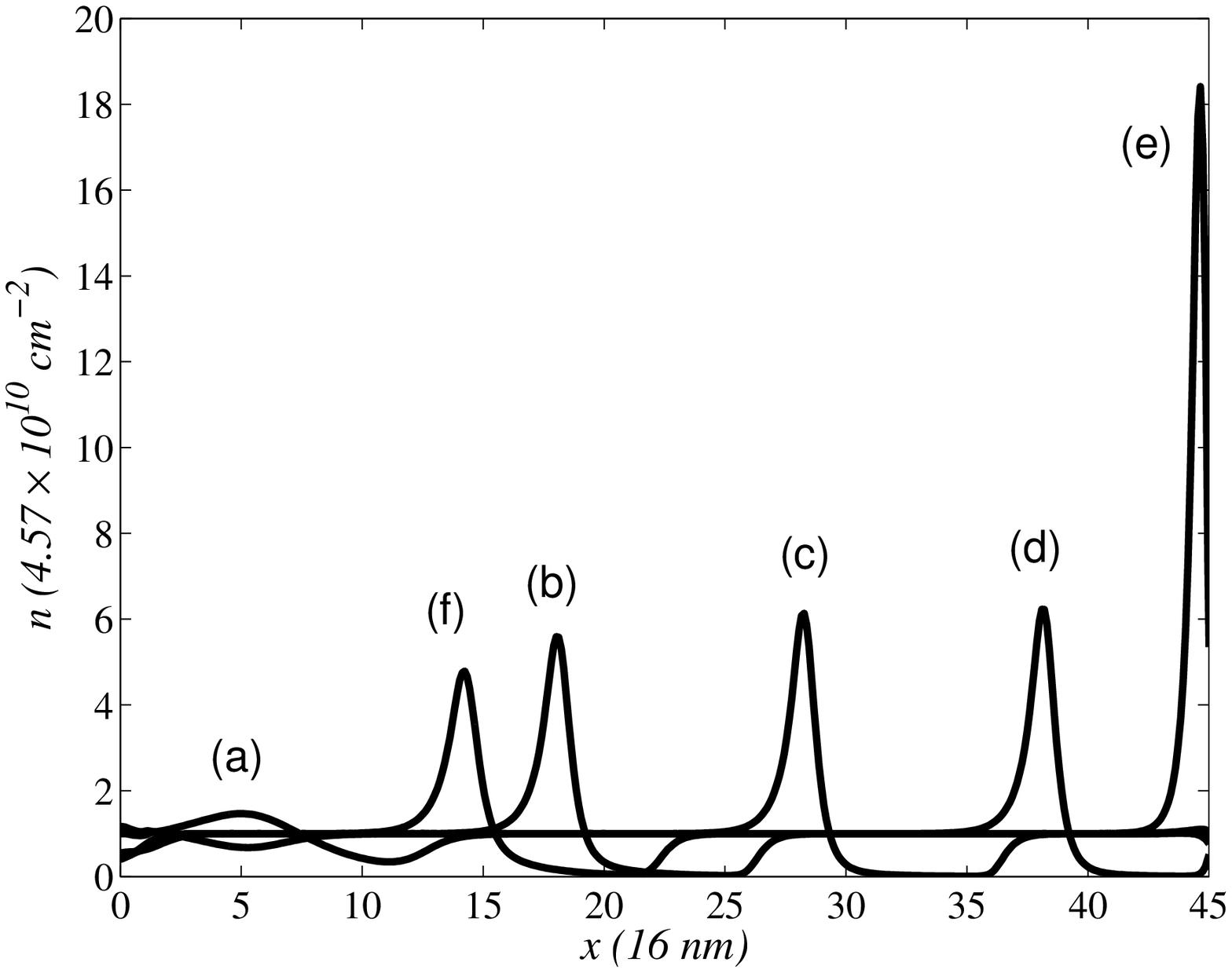} (b)
\includegraphics*[width=6cm]{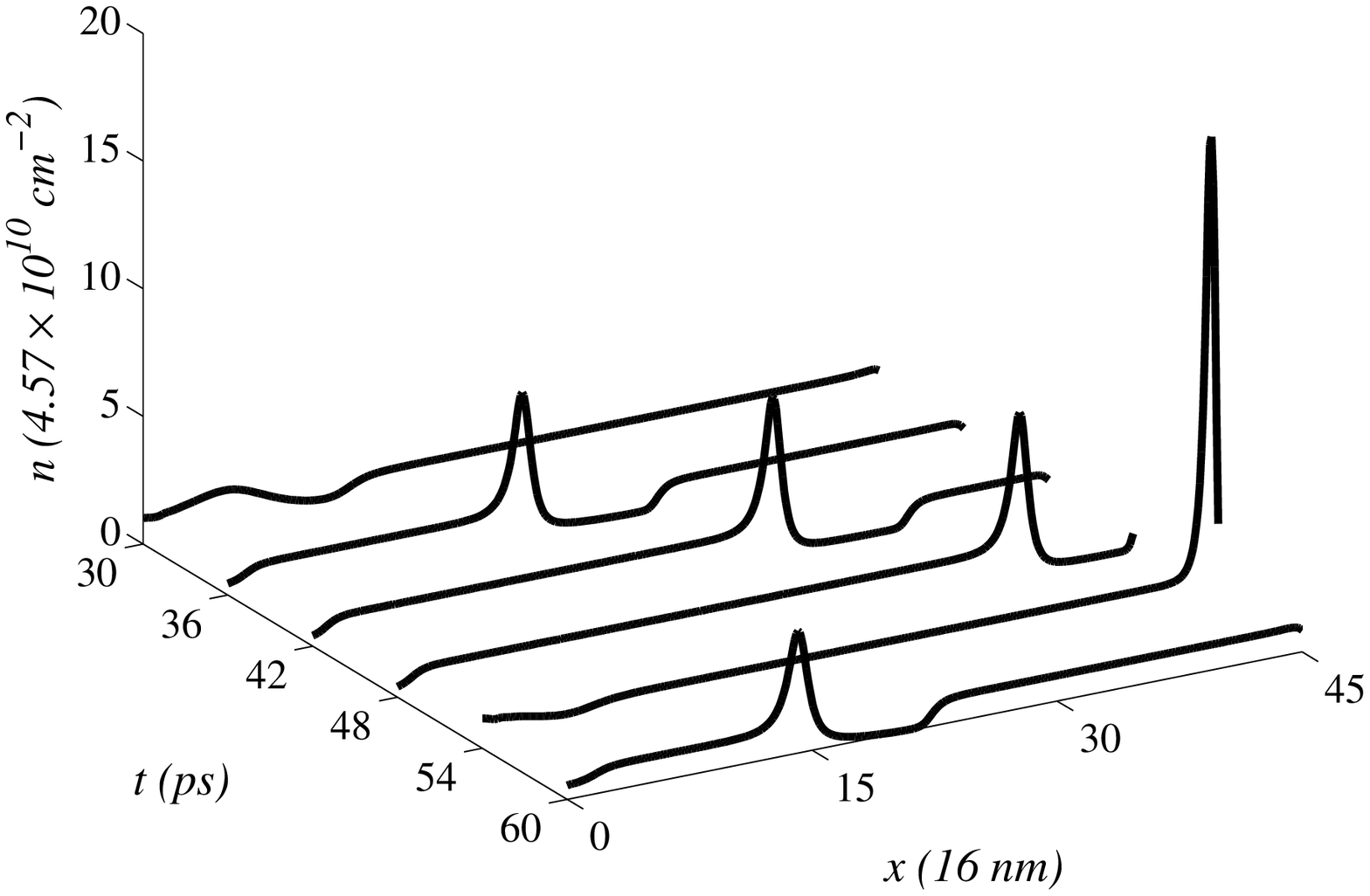} \caption{Electron
density profile for the different times as marked in Fig.\
\ref{fig1}.}\label{fig7}
\end{center}
\end{figure}

\begin{figure}
\begin{center}
\includegraphics*[width=12.5cm]{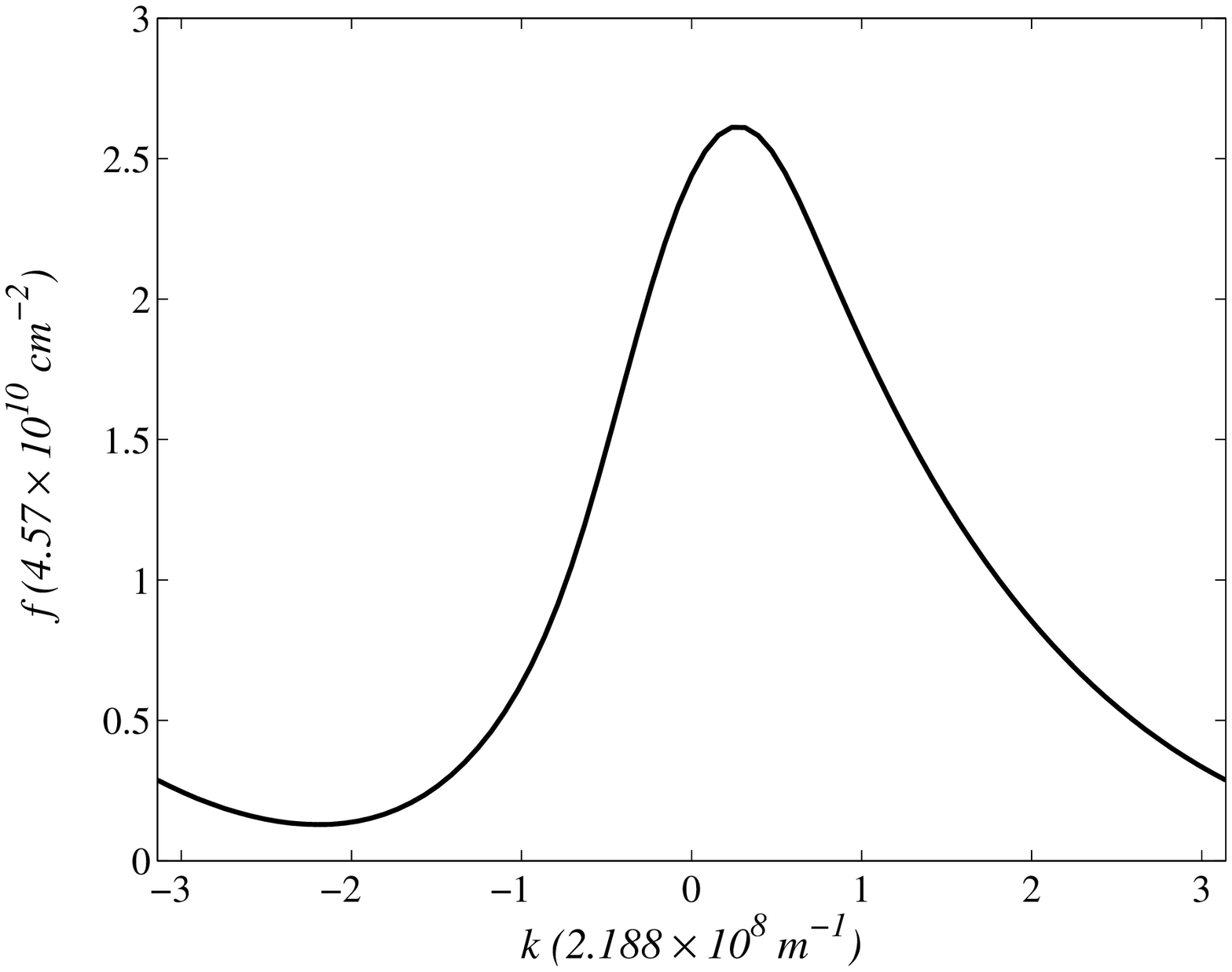}
\caption{Distribution function versus wave vector at $t=0$
ps.}\label{fig7.2}
\end{center}
\end{figure}

In Figure~\ref{fig6}, we have depicted snapshots of the distribution function $f(x,k,t)$ for
different times as marked in Fig.\ \ref{fig1} ($30$ ps, $36$ ps, $42$ ps, $48$ ps, $54$ ps,
$60$ ps) during one period of the self-oscillations. The structure of the distribution function
is shown more clearly in the density plots depicted in Fig.\ \ref{fig6.2} for the same times.
The electron density profiles at the these times are shown in Figure~\ref{fig7}. We observe
that the distribution function has a local maximum at location of the
peak of electron density. Similarly, $f$ and $n$ have local minima
at the same positions. The distribution function has a local
maximum at a positive $k$ (cf.\ Fig.\ \ref{fig6.2}), and this situation persists from the
initial time onwards; cf.\ Figure~\ref{fig7.2}.

\begin{figure}
\begin{center}
(a) \includegraphics*[width=6cm]{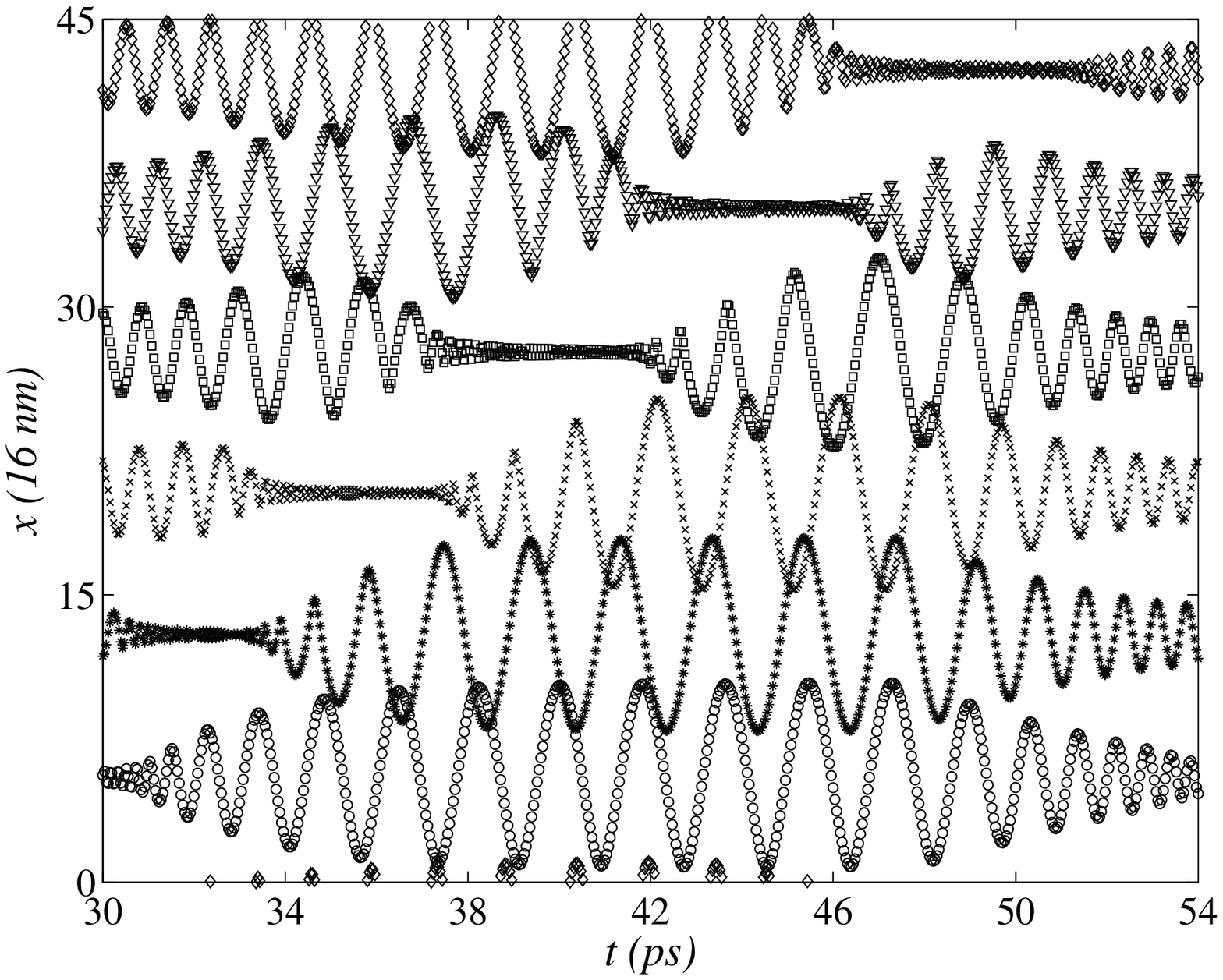} (b)
\includegraphics*[width=6cm]{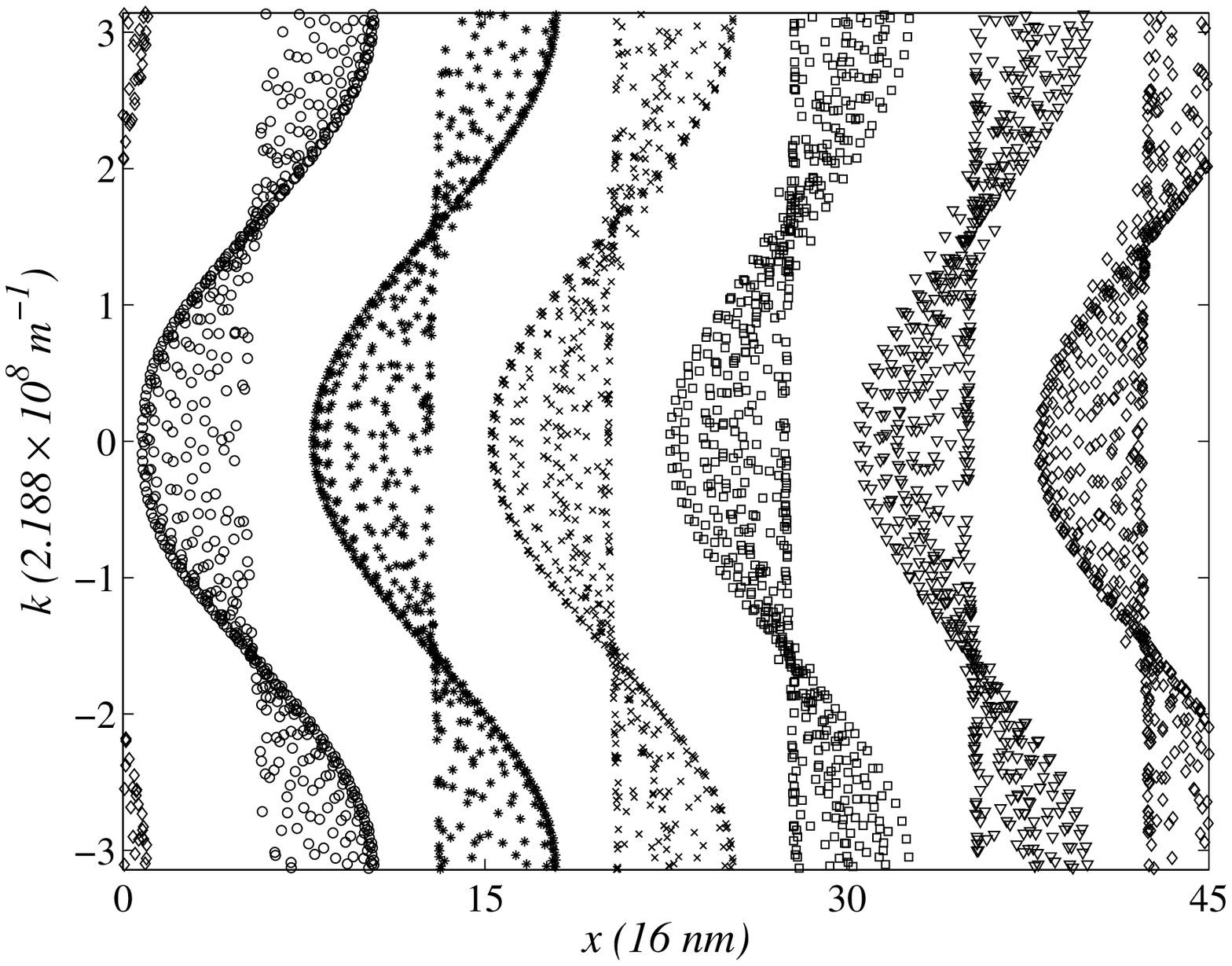} \caption{Time evolution of
the position for six particles whose initial dimensionless wave vector is $-3.126$:
(a) position vs time, (b) wave vector vs position for the same particles.}
\label{fig8}
\end{center}
\end{figure}

\begin{figure}
\begin{center}
\includegraphics*[width=12.5cm]{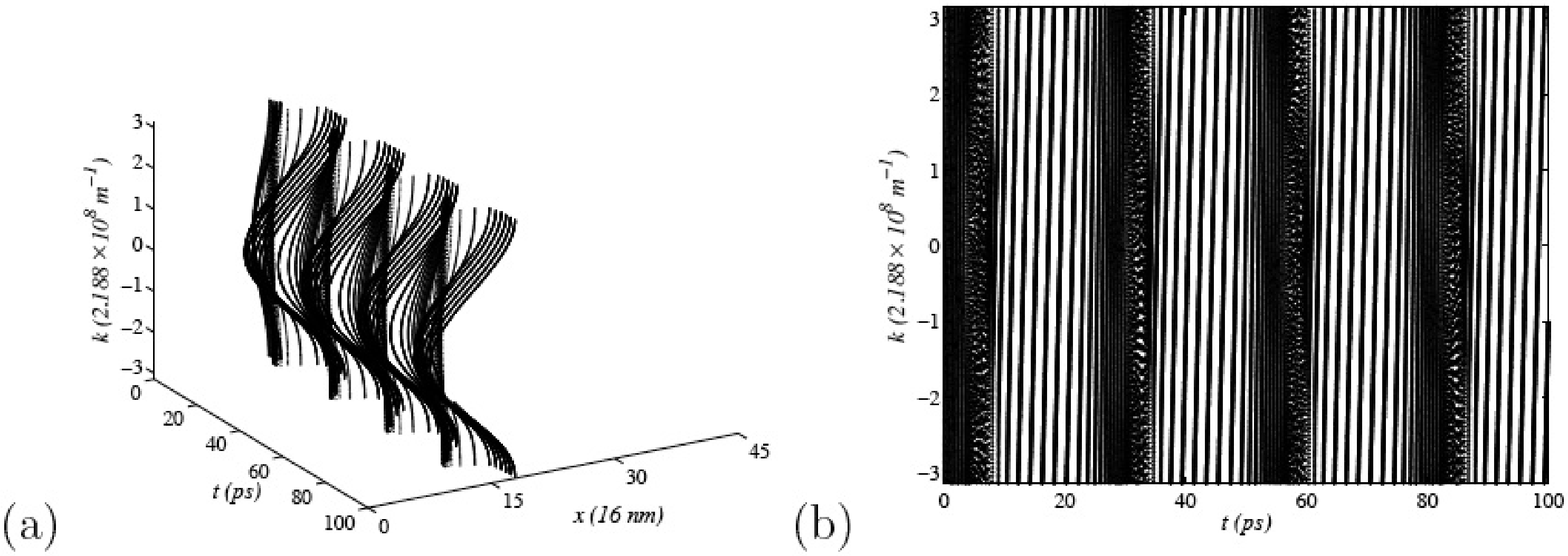} 
\caption{Time evolution of
the particle with initial dimensionless position and wave vector
$(13.986,-3.126)$ during several oscillation periods.}\label{fig9}
\end{center}
\end{figure}

\begin{figure}
\begin{center}
\includegraphics*[width=12.5cm]{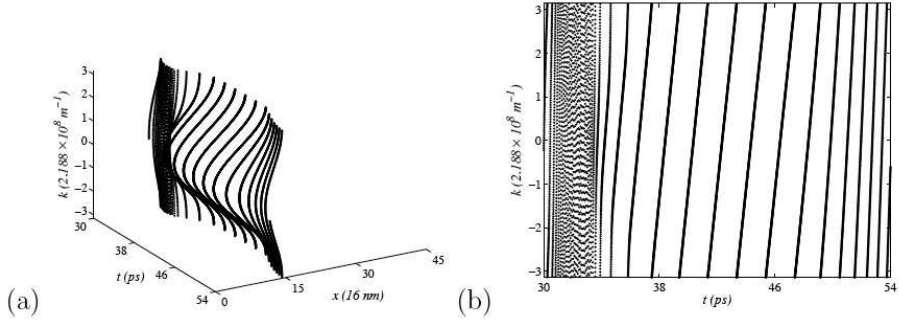} 
 \caption{Time evolution of
the particle with initial dimensionless position and wave vector
$(13.986,-3.126)$ during one oscillation period.}\label{fig9.2}
\end{center}
\end{figure}

Figure~\ref{fig8}(a) shows the time evolution of the position for
six particles, whereas Figure~\ref{fig8}(b) shows the wave vector
vs position for the same particles. The motion of the particles is
a superposition of an uniform motion and an oscillation about it.
Comparing Figure~\ref{fig8}(a) with Fig.\ \ref{fig3}(a), we
observe that the particle positions oscillate with very small
amplitudes when the electric field has a local maximum at their
locations and these amplitudes become larger once the pulse has
surpassed the particles. In contrast with these great changes in
oscillation amplitude of the particle positions, the wave vectors
of the particles oscillate with almost constant amplitudes, as
shown in Fig.\ \ref{fig8}(b), \ref{fig9} and \ref{fig9.2}.
Since the evolution of the particle wave vector is more
regular than the evolution of the particle position, we can save
mesh points on the wave vectors. Recalling that the wave vector is a
periodic variable, its boundary condition is as follows: when one
particle goes out of the domain at $k=\pi$, it is reintroduced at $k=-\pi$.
This condition can be readily observed in Figures \ref{fig9} and \ref{fig9.2}.

\section{Convergence of the method}

We have checked the convergence of the method in terms of the
number of particles $N$, mesh points $M_x$ and $M_k$,  and time step $dt$. 
Since the Fermi-Dirac weights in (\ref{eulerw}) and the electric field in (\ref{eulerk})
have to be calculated by 
interpolation over mesh points, all these parameters are important for the
convergence of the calculations over particles and of the calculations over the mesh. 
The physical parameters are the same as in the previous section.

\begin{figure}
\begin{center}
 \includegraphics*[width=12.5cm]{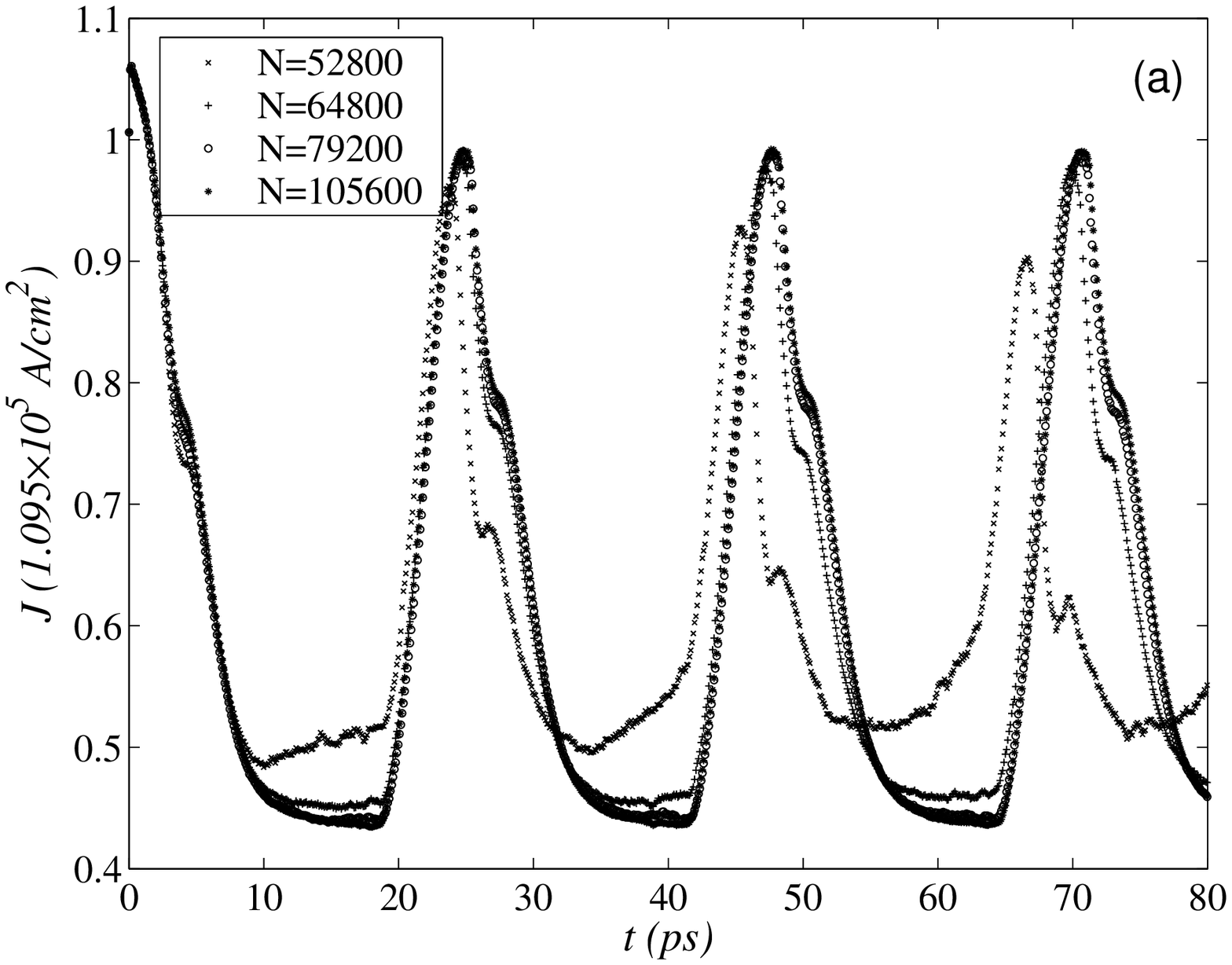}\\
\includegraphics*[width=12.5cm]{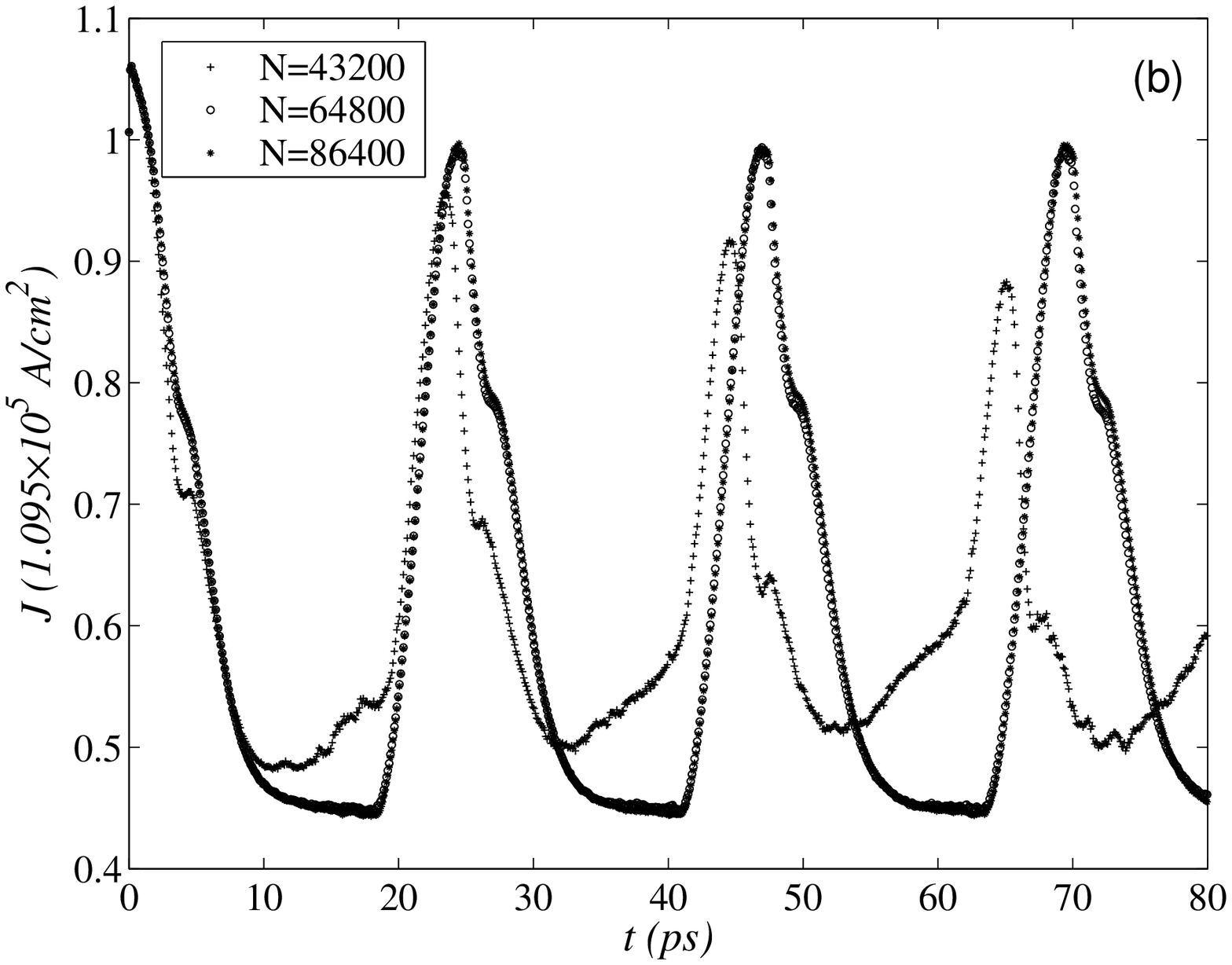} 
\caption{Current versus time for different number of particles $N$ when $M_{k}=80$
and $d t=0.008$ ps. (a) $M_{x}=440$ and (b) $M_{x}=360.$}
\label{figpart}
\end{center}
\end{figure}

Figure \ref{figpart} shows the evolution of the current density
for different number of particles $N$. We have kept the time step
and the wave number mesh points fixed at the values $M_{k}=80$ and
$dt=0.008$ ps. We observe that we need different $N$ for convergence of the
calculation depending on the value of $M_x$. In Fig.\ \ref{figpart}(a), for
$M_{x}=440$ position mesh points, we have chosen $N$ so that
$N/(M_{x}M_{k})$ takes on the values 1.5, 1.84, 2.25 and 3, whereas
in Fig.\ \ref{figpart}(b), $M_{x}=360$ and $N$ is chosen so that
$N/(M_{x}M_{k})$ takes on the values 1.5, 2.25 and 3. We observe
that for $dt=0.008$ ps and $N/(M_{x}M_{k}) \geq 2.25$ the results do not change if we
increase the number of particles. In particular, for $N=64800$,
$N/(M_{x}M_{k})=1.84$ if $M_x=440$ and $N/(M_{x}M_{k})=2.25$ if
$M_x=360$. In the first case shown in Fig.\ \ref{figpart}(a), we
need more particles for the method to converge whereas in the
second case,  Fig.\ \ref{figpart}(b) shows that we do not improve
our results by increasing $N$. The convergence range of $N/(M_{x}M_{k})$
depends slightly on the time step $dt$: if $dt=0.002$ ps, $M_{x}=440$, $M_{k}=80$,
our calculations yield indistinguishable curves $J(t)$ for $N/(M_{x}M_{k})= 3,\, 4$,
but not for $N/(M_{x}M_{k})=2.25$. Thus we have found that it is advisable to
select $N$ so that $3 \leq N/(M_{x}M_{k}) \leq 4.5$: numerical results are 
indistinguishable when $N/(M_{x}M_{k})\geq 3$ and the computational cost
is not very large if we keep $N/(M_{x}M_{k})\leq 4.5$.

\begin{figure}
\begin{center}
\includegraphics*[width=12.5cm]{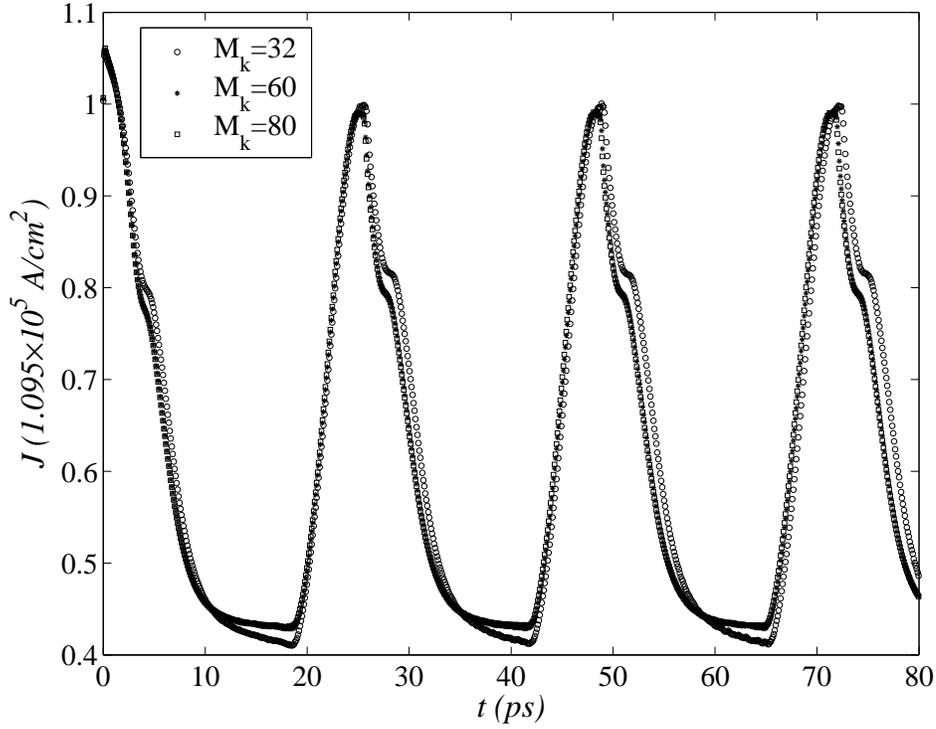}
\caption{Current versus time for different number of wave vector
mesh points when $M_{x}=520,$ $N=200000$ and $d
t=0.008$ ps.}\label{fig10}
\end{center}
\end{figure}

\begin{figure}
\begin{center}
\includegraphics*[width=12.5cm]{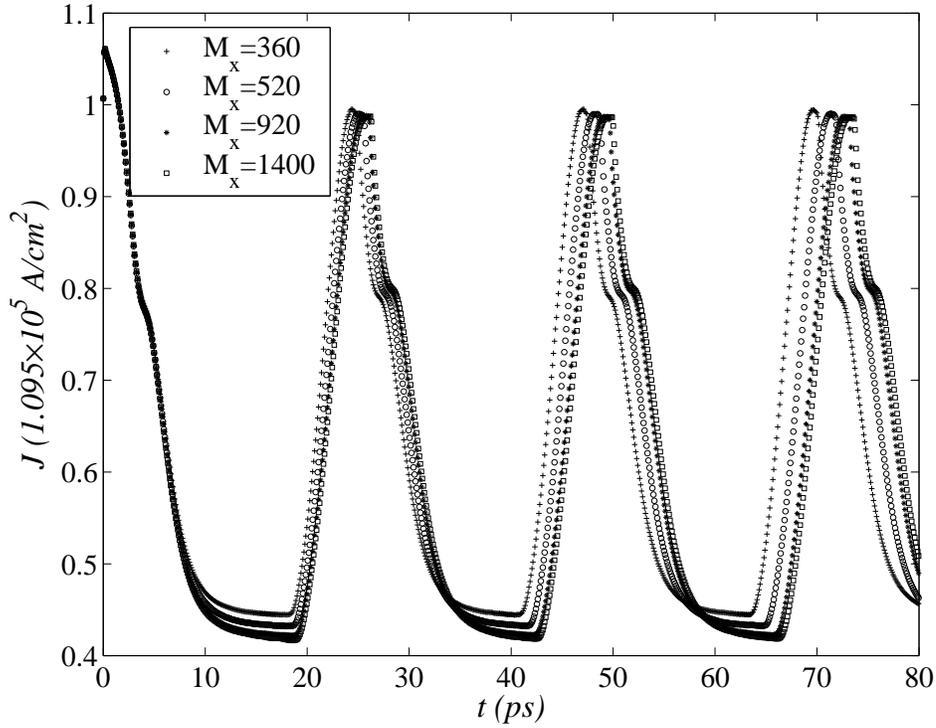}
\caption{Current versus time for different number of position mesh points when
$M_{k}=80,$ $N=480000$ and $dt=0.008$ ps.}
\label{fig11}
\end{center}
\end{figure}

Figures~\ref{fig10} and \ref{fig11} show the evolution of the
total current density in simulations having a different number of
mesh points when the time step is $dt= 0.008$ ps. In
Fig.\ \ref{fig10}, different wave vector mesh points are
considered for $M_{x}=520$ and $N=200000.$ We can check that we
do not need to have a very fine grid in $k$ because we obtain the
same results with $M_{k}= 60$ and $M_{k}=80$. In Fig.\
\ref{fig11}, different numbers of position mesh points are
considered for $M_{k}=80$ and $N=480000.$ For $M_{x}=920$ and
larger our numerical curves overlap.

\begin{figure}
\begin{center}
\includegraphics*[width=12.5cm]{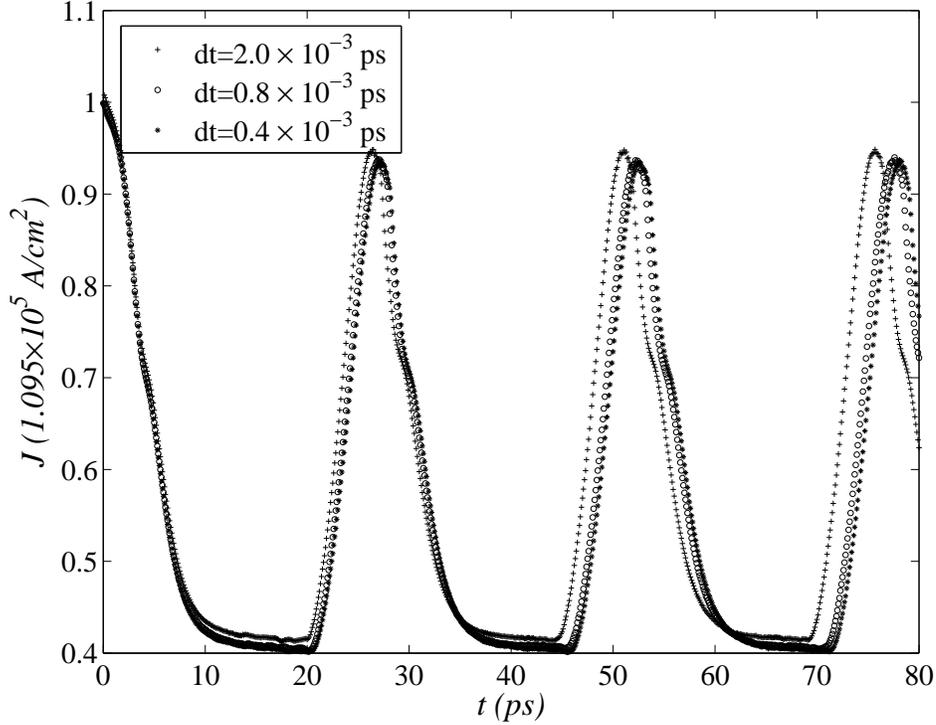}
\caption{Current versus time for different time steps when
$N=90000$, $M_{x}=260$ and $M_{k}=80$.}\label{fig12}
\end{center}
\end{figure}

Lastly, Figure \ref{fig12} shows the evolution of the total
current density for different time steps in simulations with
90000 particles and 260 mesh grid points for the position and 80
for the wave vector. We observe that our results are similar for
time steps $dt=8\times10^{-4}$ ps and smaller.

Figures \ref{fig10} to \ref{fig12} show that the shape of $J(t)$
is similar for different mesh points and time steps: the device behavior is
qualitatively correct even if we
take fewer mesh points or larger time steps than needed to attain
a numerically precise current vs time graph. Smaller $M_{k}$, $M_{x}$
and larger $dt$ result in slightly smaller oscillation periods and slightly larger
oscillation amplitudes. 

Our numerical simulations have been carried out using a 
Matlab code in a computer with a Genuine Intel(R) CPU T2050 @ 1.60GHz processor
with a 1595 MHz speed. Several computation times for time steps
$dt$ of 0.008 and 0.002 ps and 10000 time steps are shown in Table 2. Clearly, the time
the computer takes to calculate one time step $dt$ decreases as the number of particles, 
$M_x$ or $M_k$ decrease. Except for the last row in Table 2, all rows satisfy $N/(M_x M_k)
\geq 2.25$, and the corresponding particle numbers and $x$ and $k$ mesh points 
produce accurate enough results.


\[\hskip -8mm
\begin{array}{c}
\begin{array}{|c|c|c|c|c|c|}
  \hline
   N & M_x & M_k & \frac{N}{M_{x}M_{k}}& dt \mbox{ (ps)} & \mbox{(C.T.)/\# steps (seconds)}\\
   \hline
    480000 & 920 & 80 & 6.52 & 0.008 & 2.64 \\
    \hline
     480000 & 360 & 80 & 16.67 & 0.008 & 2.12 \\
    \hline
    200000 & 520 & 80 & 4.81 & 0.008 & 1.19 \\
    \hline
     200000 & 520 & 32 & 12.02 & 0.008 & 1.15 \\
  \hline
    140800 & 440 & 80 & 4.00 & 0.002 & 0.87 \\
  \hline
    105600 & 440 & 80 & 3.00 & 0.002 & 0.73 \\
  \hline
    79200 & 440 & 80 & 2.25 & 0.002 & 0.65 \\
   \hline
    52800 & 440 & 80 & 1.50 & 0.002 & 0.53 \\
    \hline
\end{array}
\\
{\rm Table \, 2: \, Computer \, time  \,  (C.T.).}
\end{array}
\]

\section{Conclusion}
We have proposed a deterministic weighted particle method to
numerically solve for the first time the semiclassical
Boltzmann-BGK-Poisson system of equations with periodic miniband
energy dispersion relation. This system describes vertical
electron transport in a GaAs/AlAs superlattice under dc voltage
bias conditions. When using appropriate values for the injecting
contact conductivity and voltage, we find a stable self-sustained
oscillation of the current through the structure which corresponds
to periodic nucleation of electric field pulses at the injecting
contact that then move to the receiving contact. The pulses have a
large electron density on their trailing edges which implies large
gradients of the electric field there. These gradients are well
resolved by particles having large weights there, which is one of
the advantages of using the weighted particle numerical method.
Our results agree with experimental observations
\cite{HGS96,BGr05} and confirm the validity of the Chapman-Enskog
perturbation method used to derive a drift-diffusion equation for
high electric fields \cite{BEP-03}. In fact, the electric field
profile and the total current density obtained by numerically
solving the the drift-diffusion equation \cite{ESC-BON-06}
agree very well with the numerical solution of the kinetic
equations obtained in the present work. Having solved the kinetic
equations directly, we can obtain the evolution of the
distribution function and its relevant moments such as electron
density, current density and average energy. The present work
paves the way to numerically solving interesting problems in
nanoelectronics and spintronics that are described by related
quantum kinetic equations with more than one miniband \cite{BBA}.

{\bf Acknowledgements:} This work has been supported by the
Ministry of Science and Innovation grants FIS2008-04921-C02-02 (EC
and AC) and FIS2008-04921-C02-01 (LLB).
%
%

\end{document}